\documentclass[a4paper,aps,prd,twocolumn,showpacs,groupedaddress,nofootinbib]{revtex4}
\usepackage{graphicx}
\usepackage{times}
\usepackage{color}

\newcommand{\ie}{\textit{i.e.,}~}
\newcommand{\eg}{\textit{e.g.,}~}

\begin{document}

\title{Collapse of differentially rotating neutron stars and
  cosmic censorship}

\author{Bruno Giacomazzo}
\affiliation{Department of Astronomy, University of Maryland, College Park, Maryland USA}
\affiliation{Gravitational Astrophysics Laboratory, NASA Goddard Space Flight Center, Greenbelt, Maryland USA}

\author{Luciano Rezzolla}
\affiliation{Max-Planck-Institut f\"ur Gravitationsphysik,
Albert Einstein Institute, Golm, Germany}
\affiliation{Department of Physics and Astronomy, Louisiana State
University, Baton Rouge, Louisiana USA }

\author{Nikolaos Stergioulas}
\affiliation{Department of Physics, Aristotle University of
Thessaloniki, Thessaloniki, Greece}

\date{\today}

\begin{abstract}
We present new results on the dynamics and gravitational-wave emission
from the collapse of differentially rotating neutron stars. We have
considered a number of polytropic stellar models having different
values of the dimensionless angular momentum $J/M^2$, where $J$ and
$M$ are the asymptotic angular momentum and mass of the star,
respectively. For neutron stars with $J/M^2<1$, \textit{i.e.},
``sub-Kerr'' models, we were able to find models that are dynamically
unstable and that collapse promptly to a rotating black hole. Both the
dynamics of the collapse and the consequent emission of gravitational
waves resemble those seen for uniformly rotating stars, although
with an overall decrease in the efficiency of gravitational-wave
emission. For stellar models with $J/M^2>1$, \textit{i.e.}
``supra-Kerr'' models, on the other hand, we were not able to find
models that are dynamically unstable and all of the computed
supra-Kerr models were found to be far from the stability
threshold. For these models a gravitational collapse is possible only
after a very severe and {artificial} reduction of the pressure, which
then leads to a torus developing nonaxisymmetric instabilities and
eventually contracting to a stable axisymmetric stellar
configuration. While this does not exclude the possibility that a
naked singularity can be produced by the collapse of a differentially
rotating star, it also suggests that cosmic censorship is not
violated and that generic conditions for a supra-Kerr progenitor do
not lead to a naked singularity.
\end{abstract}


\maketitle

%%%%%%%%%%%%%%%%%%%%%%%%%%%%%%%%%%%%%%%%%
\section{\label{sec:intro} Introduction}
%%%%%%%%%%%%%%%%%%%%%%%%%%%%%%%%%%%%%%%%%

Differentially rotating neutron stars, either dynamically stable or
unstable, can be the result of several astrophysical scenarios such as
stellar-core collapse or binary neutron star mergers (see, {\it e.g.}
Refs.~\cite{Anderson2007, Baiotti08, Anderson2008, Baiotti:2009gk,
  Etienne08, Giacomazzo:2009mp, Kiuchi2009, Abdikamalov:2009aq,
  Rezzolla:2010, Giacomazzo:2010, Rezzolla:2011} for some recent
works). Because of their differential rotation, these stars can
support masses higher than if they were uniformly
rotating~\cite{baumgarte00}. Furthermore, if $J$ and $M$ are,
respectively, the angular momentum and the gravitational mass of the
star, differentially rotating models can reach values of the
dimensionless spin parameter $J/M^2 >1$; this is not possible for
stars in uniform rotation, at least when described by a hadronic
equation of state (EOS), in which case $J/M^2 \lesssim
0.7$~\cite{friedman86}. Finally, when hypermassive (that is, having
masses larger than the one of the associated mass-shedding
configuration in uniform rotation), differentially rotating stars can
be related to events such as gamma-ray bursts (GRBs). Hypermassive
differentially rotating neutron stars can indeed be formed after the
merger of binary neutron stars and their eventual collapse can produce
a spinning black hole surrounded by a hot and massive disk, which
could generate the relativistic jets that are observed in short
GRBs~\cite{Rezzolla:2011}.

The investigation of the collapse of uniformly rotating neutron stars
to rotating black holes was first studied in detail in
Ref.~\cite{whisky}, where a specific set of dynamically unstable
models was constructed (named $\mathtt{D1}$ to $\mathtt{D4}$) for a
polytropic index $N=1$. The region of dynamical instability to
axisymmetric perturbations was probed by following the stability
criterion of Friedman, Ipser, and Sorkin~\cite{friedman88}, who
suggested that the turning point along constant angular-momentum
sequences also marks the onset of a secular instability. Models
$\mathtt{D1}$ to $\mathtt{D4}$ in Ref.~\cite{whisky} were then chosen
to be near the turning-point line, but with somewhat larger central
densities so as to ensure dynamical (and not just secular) instability
(see Table 1 and Fig. 1 of Ref.~\cite{whisky})\footnote{A recent
  investigation by Takami, Rezzolla and Yoshida~\cite{Takami:2011}
  aimed at determining the neutral point along these sequences (\ie
  where the eigenfrequency of the fundamental mode goes to zero),
  shows however that the secular stability does not coincide with the
  turning point, but must be at smaller central densities.}.

In a series of works starting with Ref.~\cite{duez04}, the study of
the collapse of differentially rotating neutron stars was initiated,
both in axisymmetry and in three spatial dimensions (3D). In
Ref.~\cite{duez04}, in particular, three different models were
considered, two of which were ``sub-Kerr'', \ie with $J/M^2<1$ and one
was instead ``supra-Kerr'', \ie with $J/M^2>1$. The collapse of the
latter was obtained by artificially depleting the pressure by 99\%,
but no investigation was made on the stability of the progenitor
models. The results presented in~\cite{duez04} were based on the use
of a nonconservative numerical scheme and of a polytropic EOS
preventing the formation of strong shocks, which cannot be handled
well by nonconservative numerical methods. In a subsequent
work~\cite{duez04b}, the collapse of similar models was studied in the
presence of viscosity and the first results were published, but only
in axisymmetry, for the collapse of differentially rotating neutron
stars in the presence of a poloidal magnetic field~\cite{duez05,
  duez06a, duez06b, duez06c, stephens07}.

The main goal of this work is to reconsider the effect of differential
rotation on the collapse of rotating neutron stars and to present a
systematic investigation of the stability properties of differentially
rotating models. Special attention has been paid to the following two
questions. (1) Do stellar models exist with $J/M^2 > 1$ and that are
dynamically unstable to the collapse to black hole?  (2) If a stellar
model with $J/M^2 > 1$ is induced to collapse, does it lead to a naked
singularity, thus violating cosmic censorship?

Overall, our results can be summarized as follows:
\begin{itemize}

\item[] \textit{(i)} differentially rotating, sub-Kerr neutron-star
  models can be found that are dynamically unstable to the
  gravitational collapse to rotating black holes;

\item[] \textit{(ii)} the dynamics of sub-Kerr models is very similar to
  the one of uniformly rotating models: nonaxisymmetric instabilities
  do not have sufficient time to grow and the stars collapse promptly
  to black holes;

\item[] \textit{(iii)} the efficiency in the emission of gravitational
  radiation is comparable to that of uniformly rotating stars,
  although generically smaller since the collapse is generally slower;

\item[] \textit{(iv)} differentially rotating, supra-Kerr models that
  are dynamically unstable to the gravitational collapse to rotating
  black holes could not be found; rather, all of the supra-Kerr models
  studied were found to be dynamically stable;

\item[] \textit{(v)} a supra-Kerr model can be induced to collapse
  only through a very severe depletion of the pressure support which,
  however, does not lead to the prompt formation of a rotating black
  hole;

\item[] \textit{(vi)} because of the development of nonaxisymmetric
  instabilities, the gravitational-wave emission from supra-Kerr
  models could in principle be considerably larger than that of
  sub-Kerr models if ever induced to collapse.

\end{itemize}

The paper is organized as follows. In Sec.~\ref{sec:equations} we
briefly summarize the equations solved and the numerical
infrastructure used. In Sec.~\ref{sec:models} we present our study on
the stability of differentially rotating neutron stars with different
degrees of differential rotation and different polytropic EOSs. In
Sec.~\ref{sec:dynamics} we present the dynamics of the collapse of
four models, three with $J/M^2<1$ and one with $J/M^2>1$, and in
Sec.~\ref{sec:waves} their gravitational-wave signal. Finally, in
Sec.~\ref{sec:diffot_conclusions} we summarize and conclude.

Throughout this paper we use a spacelike signature of $(-,+,+,+)$ and a
system of units in which $c=G=M_\odot =1$. Greek indices are taken to
run from 0 to 3, Latin indices from 1 to 3 and we adopt the standard
convention for the summation over repeated indices.

%%%%%%%%%%%%%%%%%%%%%%%%%%%%%%%%%%%%%%%%%%%%%%%%%%%%%%%%%%%%%%%%%%%
\section{\label{sec:equations}Basic equations and numerical setup}
%%%%%%%%%%%%%%%%%%%%%%%%%%%%%%%%%%%%%%%%%%%%%%%%%%%%%%%%%%%%%%%%%%%

All the simulations presented here were done using the {\tt Whisky}
code which solves the general-relativistic hydrodynamic equations on
a three-dimensional numerical grid with Cartesian
coordinates~\cite{Baiotti03a}. The code has been constructed within
the framework of the {\tt Cactus} Computational
Toolkit~\cite{Goodale02a}, which provides high-level facilities such
as parallelization, input/output, portability on different platforms
and several evolution schemes to solve general systems of partial
differential equations. Clearly, special attention is dedicated to the
solution of the Einstein equations, whose matter terms in nonvacuum
spacetimes are handled by the {\tt Whisky} code.

In other words, while the {\tt Cactus} code provides at each time step
and on a spatial hypersurface the solution of the Einstein equations
\begin{equation}
\label{efes}
G_{\mu \nu}=8\pi T_{\mu \nu}\ , 
\end{equation}
where $G_{\mu \nu}$ is the Einstein tensor and $T_{\mu \nu}$ is the
stress-energy tensor, the {\tt Whisky} code provides the time evolution
of the hydrodynamic equations, expressed through the conservation
equations for the stress-energy tensor $T^{\mu\nu}$ and for the matter
current density $J^\mu$,
\begin{equation}
\label{hydro eqs}
\nabla_\mu T^{\mu\nu} = 0\;,\;\;\;\;\;\;\;\;\;\;\;\;
\nabla_\mu J^\mu = 0\;.
\end{equation}
This system of equations is then closed by an EOS which relates the
pressure to the rest-mass density and to the specific internal energy.

In what follows, and mostly for the sake of completeness, we give a
brief overview of how both the right and the left-hand side of
Eq. (\ref{efes}) are computed within the coupled {\tt
  Cactus/Whisky} codes. The equations presented have already been
discussed in several different publications, e.g. in
~\cite{Alcubierre99d,whisky,baiotti05} and we refer the interested
reader to these works for more details. We note that the
\texttt{Whisky} code can also solve the equations of
general-relativistic magnetohydrodynamics (GRMHD) within the ideal-MHD
limit~\cite{giacomazzo07,Giacomazzo:2010,Rezzolla:2011}. Hereafter,
however, in order to build the necessary understanding of the dynamics
of gravitational collapse in the presence of differential rotation, we
will consider unmagnetized fluids only, leaving the inclusion of
magnetic fields to a future study.

%-------------------------------------------
\subsection{\label{feqs}Evolution of the field equations}   
%-------------------------------------------

We use the conformal and traceless decomposition of the
Arnowitt-Deser-Misner (ADM) formulation~\cite{Arnowitt62} of the
Einstein equations as first presented in 3D in Ref.~\cite{Nakamura87},
which is based on the ADM construction and has been further developed
in~\cite{Shibata95}.  Details of our particular implementation of the
conformal traceless reformulation of the ADM system as proposed
by~\cite{Nakamura87,Shibata95,Baumgarte99} are extensively described
in~\cite{Alcubierre99d,Alcubierre02a} and will not be repeated here.

%%-------------------------------------------
%\subsubsection{Gauge choices}
%%-------------------------------------------

The code is designed to handle arbitrary shift and lapse conditions,
which can be chosen as appropriate for a given spacetime simulation.
More information about the possible families of spacetime slicings
which have been tested and used with the present code can be found
in~\cite{Alcubierre99d,Alcubierre01a}. Here, we limit ourselves to
recalling details about the specific foliations used in the present
evolutions. In particular, we have used hyperbolic $K$-driver slicing
conditions of the form
\begin{equation}
\partial_t \alpha = - f(\alpha) \;
\alpha^2 (K-K_0),
\label{eq:BMslicing}
\end{equation}
with $f(\alpha)>0$ and $K_0 \equiv K(t=0)$, with $K$ being the trace
of the extrinsic curvature and $\alpha$ the lapse function.  All the
simulations discussed in this paper were performed using condition
(\ref{eq:BMslicing}) with $f=2/\alpha$. For the spatial gauge we use
one of the ``Gamma-driver'' shift conditions proposed
in~\cite{Alcubierre01a} (see also \cite{Alcubierre02a}).
In particular, all the results reported here have been obtained using
the hyperbolic Gamma-driver condition,
\begin{equation}
\partial^2_t \beta^i = F \, \partial_t \tilde\Gamma^i - \eta \,
\partial_t \beta^i,
\label{eq:hyperbolicGammadriver}
\end{equation}
where $\beta^i$ is the shift, $\tilde{\Gamma}^{i}$ are the ``conformal
connection functions'' and $F$ and $\eta$ are, in general, positive
functions of space and time. For the hyperbolic Gamma-driver
conditions it is crucial to add a dissipation term with coefficient
$\eta$ to avoid strong oscillations in the shift. Experience has shown
that by tuning the value of this coefficient it is possible to almost
freeze the evolution of the system at late times. We typically choose
$F={3}/{4}$ and $\eta=3$ and do not vary them in time.

The singularity-avoiding properties of the above gauge choices have
proved equally good both when using excision, as it was done in
Refs.~\cite{whisky} and~\cite{baiotti05}, and when not using
excision~\cite{baiotti06}. In particular in this paper we employ the
``no-excision'' technique introduced in Ref.~\cite{baiotti06} and we
add an artificial dissipation of the Kreiss-Oliger
type~\cite{Kreiss73} on the right-hand sides of the evolution
equations for the field variables (no dissipation is introduced for
the hydrodynamical variables). As first pointed out in
Ref.~\cite{baiotti06}, in fact, renouncing to excision and using
instead suitable ``singularity-avoiding'' slicing conditions improves
dramatically the long-term stability of the simulations, allowing for
the calculation of the gravitational waveforms well beyond the
quasi-normal-mode ringing.

%-------------------------------------------
\subsection{Evolution of the hydrodynamic equations}    
%-------------------------------------------

An important feature of the {\tt Whisky} code is the implementation of
a \textit{conservative formulation} of the hydrodynamic equations
\cite{Banyuls97}, in which the set of Eqs. (\ref{hydro eqs}) is
written in a hyperbolic, first-order, and flux-conservative form of the
type
\begin{equation}
\label{eq:consform1}
\partial_t {\mathbf q} + 
        \partial_i {\mathbf f}^{(i)} ({\mathbf q}) = 
        {\mathbf s} ({\mathbf q})\ ,
\end{equation}
where ${\mathbf f}^{(i)} ({\mathbf q})$ and ${\mathbf s}({\mathbf q})$
are the flux vectors and source terms, respectively~\cite{Font03}.  Note
that the right-hand side (the source terms) does not depend on derivatives of 
the stress-energy tensor.

An important feature of this formulation is that it allows for the
extension to a general-relativistic context of the powerful numerical
methods developed in classical hydrodynamics, in particular,
high-resolution shock-capturing schemes based on
exact~\cite{Marti99,Rezzolla01,Rezzolla03} or approximate Riemann
solvers (see Ref.~\cite{Font03} for a detailed bibliography). Such
schemes are essential for a correct representation of shocks, whose
presence is expected in several astrophysical scenarios.

For all the results presented here, we have solved the hydrodynamic
equations employing the Marquina flux formula and a third-order
PPM~\cite{Colella84} reconstruction. A third-order Runge-Kutta scheme
was then used for the evolution.

%-------------------------------------------
\subsection{\label{sec:mesh_refinement}Mesh Refinement}    
%-------------------------------------------

We solve both the {spacetime} and hydrodynamic equations on
nonuniform grids using a ``box-in-box'' mesh refinement strategy
implemented in \texttt{Whisky} via the {\tt Carpet}
driver~\cite{Schnetter-etal-03b}. This introduces two important
advantages: first, it reduces the influence of inaccurate boundary
conditions at the outer boundaries, which can be moved far from the
central source; second, it allows for the wave zone to be included
in the computational domain and thus for the extraction of important
information about the gravitational-wave emission produced during the
collapse. In practice, we have adopted a Berger-Oliger prescription
for the refinement of meshes on different levels~\cite{Berger84} and
used the numerical infrastructure described
in~\cite{Schnetter-etal-03b}.

%-------------------------------------------
\subsection{\label{gwmethod} Gravitational-wave extraction}
%-------------------------------------------
While several different methods are possible for the extraction of the
gravitational-radiation content in numerical spacetimes, we have
adopted a gauge-invariant approach in which the spacetime is matched
with the nonspherical perturbations of a Schwarzschild black hole
(see Refs.~\cite{Rupright98,cs99} for applications to Cartesian
coordinates). In practice, a set of ``observers'' is placed on
2-spheres of fixed coordinate radius $r_{\rm ex}$, where the
gauge-invariant odd-parity $Q^{({\rm o})}_{\ell m}$ and even-parity $\Psi^{({\rm
    e})}_{\ell m}$ metric perturbations~\cite{moncrief74,nagar05,nagar06}
are extracted.
From these quantities it is possible to compute the even- and
odd-parity perturbations
\begin{eqnarray}
Q^{+}_{\ell m} &=& \lambda\Psi^{({\rm e})}_{\ell m} \,,\\
Q^{\times}_{\ell m} &=& \lambda Q^{({\rm o})}_{\ell m}\,, 
\end{eqnarray}
where $\lambda \equiv \sqrt{{2(\ell+2)!} / {(\ell-2)!}}$. Using these
quantities it is also possible to compute the gravitational-wave
amplitudes in the two polarizations $h_{+}$ and $h_{\times}$ as
\begin{equation}
h_{+}-{\mathrm i}h_{\times}=\frac{1}{2r}\sum_{\ell,m}
        \left(Q^{+}_{\ell m}-{\mathrm i}
        \int_{-\infty}^{t}\!\!\!\!\!\!
        Q^{\times}_{\ell m}(t')dt'
        \right)\,_{_{-2}}Y^{\ell m}, 
\end{equation}
where $_{_{-2}}Y^{\ell m}$ are the $s=-2$ spin-weighted spherical
harmonics. We note that the approach discussed above cannot be
employed when simulating the supra-Kerr model, for which a very high
resolution is necessary and hence the use of a computational domain
with rather close boundaries. As we will comment later on in
Sec.~\ref{ssec:suprakerr_waves}, for that model the
gravitational-wave emission will be computed through the quadrupole
formula.

%%%%%%%%%%%%%%%%%%%%%%%%%%%%%%%%%%%%%%%%%%%%%%%%%%%
\section{\label{sec:models}Initial stellar models}
%%%%%%%%%%%%%%%%%%%%%%%%%%%%%%%%%%%%%%%%%%%%%%%%%%%

%-------------------------------------------
\subsection{\label{EOS_DRL}Equation of state and differential-rotation law}
%-------------------------------------------
We construct our initial stellar models using the numerical code {\tt rns} 
\cite{Stergioulas1995} (see \cite{Nozawa1998,Stergioulas2003} for evaluations 
of its accuracy) as isentropic, differentially
rotating relativistic polytropes, satisfying the polytropic EOS
\begin{equation}
p=K \rho^\Gamma,
\end{equation}
\begin{equation}
e=\rho+\frac{p}{\Gamma-1},
\end{equation}
where $p$, $\rho$, and $e$ are pressure, rest-mass density, and energy
density, respectively, while $K=100$ and $\Gamma$ are the polytropic
constant and the polytropic exponent, respectively. In our discussion
we will also use the polytropic index $N$, defined through the
relation $\Gamma \equiv 1+1/N$.

We further assume that the equilibrium models are stationary and 
axisymmetric, so that the spacetime geometry is described by a 
metric of the form
\begin{equation}
{\rm d}s^2 = -{\rm e}^{2 \nu} {\rm d}t^2 + {\rm e}^{2 \psi} ({\rm d} \phi - 
	\omega {\rm d}t)^2 + {\rm e}^{2
	\mu} ({\rm d}r^2+r^2 {\rm d} \theta^2),
\label{e:metric}
\end{equation}
where $\nu$, $\psi$, $\mu$ and $\omega$ are functions of the
quasi-isotropic coordinates $r$ and $\theta$ only. The degree of
differential rotation, as well as its variation within the star, are
essentially unknown and because of this we here employ the well-known
``{\it j}-constant'' law of differential rotation~\cite{KEH89}
\begin{equation}
A^2(\Omega_{\rm c}-\Omega) = \frac{(\Omega-\omega){\rm e}^{2\psi}}
	{1-(\Omega -\omega) {\rm e}^{2\psi}},
\end{equation}
where $\Omega\equiv\Omega(r,\theta)$ is the angular velocity,
$\Omega_{\rm c}$ is the angular velocity at the center of the star, and
$A$ is a constant (with the dimensions of length) that represents the
lengthscale over which the angular velocity changes
(see~\cite{Galeazzi:2011} for a new and more realistic law of
differential rotation). In the remainder of this Section, we will
measure the degree of differential rotation by the rescaled quantity
$\hat{A}\equiv A/r_e$, where $r_e$ is the equatorial coordinate radius
of the star. For $\hat{A}\rightarrow \infty$ uniform rotation is
recovered while a low value of $\hat{A}$ indicates a high degree of
differential rotation.

%%%%%%%%%%%%%%%%%%%%%%%%%%%%%%%%%%%%%%%%%%%%%%%%%%%%%%%%%%%
\subsection{Equilibrium models}
%%%%%%%%%%%%%%%%%%%%%%%%%%%%%%%%%%%%%%%%%%%%%%%%%%%%%%%%%%%

As mentioned in the introduction, when studying the gravitational
collapse of a rotating neutron star to a Kerr black hole, an
interesting question is about what happens to a configuration with an
initial dimensionless spin parameter $J/M^2>1$ (\textit{i.e.}, a
supra-Kerr model). If the cosmic censorship conjecture is expected to
hold, such models cannot collapse promptly to Kerr black holes, which
are limited to $J/M^2 \leq 1$. Rather, one expects that the transition
to a black hole, when it occurs, takes place after the stellar model
has shed some of its angular momentum. This was indeed what was shown
to happen in the previous study of Ref.~\cite{duez04}, where a
dynamically stable supra-Kerr model was induced to collapse after a
dramatic depletion of the pressure support.

Here, we consider again this interesting question but, before discussing
in more detail the features of the gravitational collapse of either
supra- or sub-Kerr models, it is worth discussing the properties of the
equilibrium models that can be computed with the prescriptions of the EOS
and of the differential-rotation law discussed in the previous
section~\ref{EOS_DRL}. As will become apparent, the study of these
equilibrium models will be quite revealing for the stability properties
and hence for what is realistic to expect from the gravitational collapse
of differentially rotating neutron stars. To this end, we have
constructed a large set of initial models for various values of the
polytropic index $N$ and degree of differential rotation $\hat{A}$,
reaching close to the mass-shedding limit and spanning a wide range of
central densities.

\begin{figure}
  \begin{center}
  \includegraphics[width=0.45\textwidth]{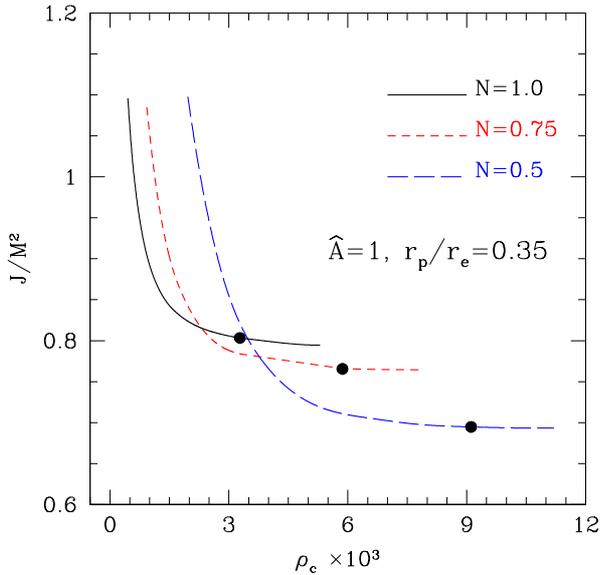}
  \end{center}
  \caption{\label{fig:max_JoM2_rho_c}$J/M^2$ as a function of central
    rest-mass density $\rho_c$ for models with polytropic indices
    $N=0.5$, $N=0.75$ and $N=1.0$, when the rotation law and the
    polar-to-equatorial coordinate axis ratio are fixed to
    $\hat{A}=1.0$ and $r_p/r_e=0.35$, respectively. The circle denotes
    roughly the separation between stable (at the left of the circle)
    and unstable (at the right) models along each sequence.}
\end{figure}

Figure~\ref{fig:max_JoM2_rho_c} shows the value of $J/M^2$ as a
function of central rest-mass density $\rho_c$ for the three different
EOSs with $N=0.5$, $N=0.75$, and $N=1.0$. In these sequences the
rotation law and the polar-to-equatorial coordinate axis ratio are
fixed to $\hat{A}=1.0$ and $r_p/r_e=0.35$, respectively. The choice of
$\hat{A}=1.0$ is a typical one, representing moderate differential
rotation (the angular velocity at the axis and at the equator differ
by a factor of $\sim 3$), while the axis ratio of $0.35$ refers to
very rapidly rotating models near the mass-shedding limit (when the
limit exits).  Along each sequence, we mark by a filled circle the
model which roughly separates stable models (at lower central
rest-mass densities) from unstable models (at higher central rest-mass
densities).  As we do not know precisely which are the marginally
stable models (no simple turning-point criterion exists in the case of
differential rotation), we use as a reference the stability limit of
the nonrotating models and thus we mark with a circle the central
rest-mass density of the nonrotating model having the maximum mass for
each EOS\footnote{We also note that the value of the central rest-mass
  density separating stable from unstable models in the case of
  uniformly rotating neutron stars changes by less than $\approx 5 \%$
  when moving from a nonrotating sequence to a maximally rotating
  one. As a result, using a nonrotating model to mark the stability is
  a very good approximation which does not affect the results
  presented here.}. Stated differently, all models to the right of the
circles are expected to be dynamically unstable or at least very close
to the instability threshold.

As becomes clear from this figure, all \textit{unstable models} we
were able to construct have $J/M^2<1$, \textit{i.e.}, \textit{are
  sub-Kerr}. In contrast, in order to find supra-Kerr models, one must
reach very low central rest-mass densities, where equilibrium models
are instead expected to be (very) stable against axisymmetric
perturbations. Interestingly, for the particular sequences considered
here, the value of $J/M^2$ in the unstable region (\textit{i.e.}, to
the right of the filled circles) becomes nearly constant for all the
considered values of the polytropic indices considered; this
represents an additional evidence that all unstable models are indeed
sub-Kerr.

\begin{figure}
  \begin{center}
  \includegraphics[width=0.45\textwidth]{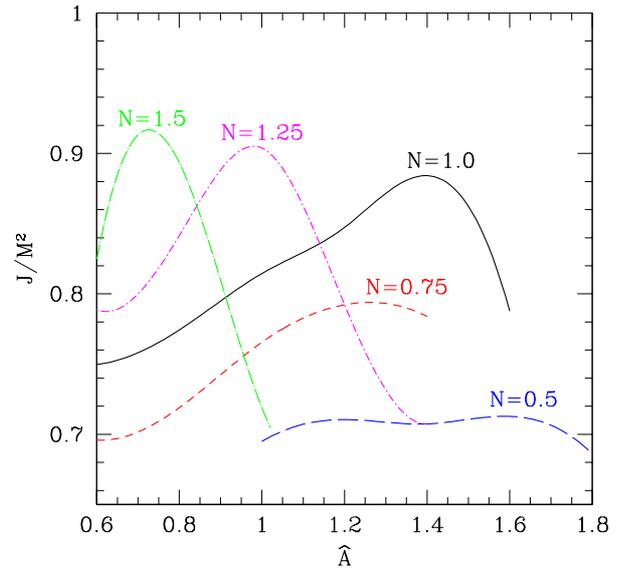}
  \end{center}
  \caption{\label{fig:max_JoM2_A}$J/M^2$ of the most rapidly rotating
    models with a central rest-mass density equal to that of the
    maximum-mass nonrotating models (\textit{i.e.}, for the models
    marked with the circles in Fig.~\ref{fig:max_JoM2_rho_c}) as a
    function of the rotation law parameter $\hat{A}$ and for different
    values of $N$. All models have $J/M^2<1$ indicating the difficulty
    of constructing unstable supra-Kerr models.}
\end{figure}

In order to investigate further the effect of the
differential-rotation-law parameter $\hat{A}$ and of the EOS on the
above conclusion, we have investigated a large number of rapidly
rotating models, spanning a wide range of values for $\hat{A}$
(between $0.6$ and $1.8$) and a wide range of polytropic indices
(between $0.5$ and $1.5$).  In all cases, we have computed the value
of $J/M^2$ of the most rapidly rotating models we could construct
(which was normally close to the mass-shedding limit, when it exists)
for a central rest-mass density equal to that of the maximum-mass
nonrotating model (\textit{i.e.}, for the models marked by circles in
Fig.~\ref{fig:max_JoM2_rho_c}). The results of this analysis are
collected in Fig.~\ref{fig:max_JoM2_A}, which shows that all the
models with a central density equal to the maximum-mass nonrotating
stars have $J/M^2<1$, when $N$ and $\hat{A}$ are allowed to vary. This
remains true also when considering the unstable models in
Fig.~\ref{fig:max_JoM2_rho_c} ({\it i.e.}, the ones with an higher
central density), which all have a value of $J/M^2$ lower than the
models shown in Fig.~\ref{fig:max_JoM2_A}. It is therefore evident
that no combination of $N$ and $\hat{A}$ could yield an unstable
supra-Kerr model. All of these results provide strong evidence that
all \textit{supra-Kerr models} examined here \textit{are stable.}

Bearing this in mind, it should be noted that because our numerical
method does not reach exactly the mass-shedding limit for any degree
of differential rotation (it is difficult to achieve convergence at
very small values of the axis ratio $r_p/r_e$) and since the existence
of a bifurcation between quasispheroidal and quasitoroidal
models\footnote{We define as ``quasi-spheroidal'' those models having
  the central and maximum rest-mass density being coincident, while we
  define as ``quasi-toroidal'' those models having the maximum of the
  rest-mass density not located at the center of the star.} with the
same axis ratio and central density has not been investigated yet, we
cannot strictly exclude the existence of supra-Kerr unstable models.

\begin{figure}
  \begin{center}
  \includegraphics[width=0.45\textwidth]{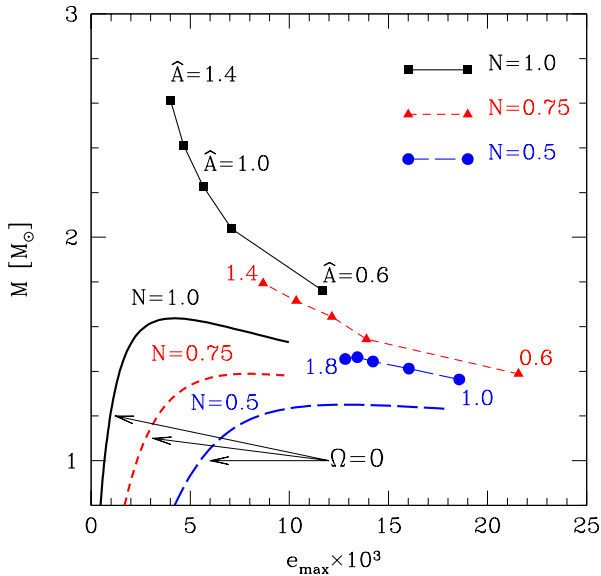}
  \end{center}
  \caption{\label{fig:max_JoM2_emax}The solid, short-dashed and
    long-dashed lines (upper curves with individual models marked)
    represent the gravitational mass $M$ of some of the unstable
    models shown in Fig.~\ref{fig:max_JoM2_A}, as a function of the
    maximum energy density, for $N=0.5,0.75$ and $1.0$. The different
    values of $\hat{A}$ are reported near each model. The lower curves
    show the corresponding nonrotating sequence of models for each of
    the above EOSs.}
\end{figure}

The rapidly rotating models with $N=1.0$, $N=0.75$ and $N=0.5$ shown
in Fig.~\ref{fig:max_JoM2_rho_c} are also shown in
Fig.~\ref{fig:max_JoM2_emax} in a mass vs maximum energy density
plot. Since the most rapidly rotating models with differential
rotation and small axis ratio are quasitoroidal, the corresponding
maximum energy density is larger than the central energy density by a
factor of roughly 2, depending on the degree of differential
rotation. It is not yet known whether the value of the central energy
density or of the off-center maximum energy density is more important
in determining the stability to axisymmetric perturbations of
quasitoroidal models. Therefore, the models shown in
Figs.~\ref{fig:max_JoM2_A} and \ref{fig:max_JoM2_emax} could either
be only marginally stable or unstable or strongly unstable.
Nevertheless, the fact that the central density of models in
Fig.~\ref{fig:max_JoM2_rho_c} with $J/M^2>1$ is at least a factor of
3 smaller than the central density of the corresponding
maximum-mass nonrotating models, indicates that even if all models in
Fig.~\ref{fig:max_JoM2_emax} are well inside the dynamically unstable
region, there should still be no supra-Kerr unstable models for the
parameter range examined.

 \begin{table*}
 \caption{\label{tab:diffrot_id}Initial data for the different stellar
   models. The different columns refer, respectively, to the central
   rest-mass density $\rho_c$ and its maximum $\rho_{\mathrm{max}}$,
   the ratio of the polar to the equatorial coordinate radii
   $r_p/r_e$, the total gravitational mass $M$, the circumferential
   equatorial radius $R_e$, the central angular velocity $\Omega_c$,
   the ratio of rotational kinetic energy to gravitational binding
   energy $T/|W|$, the ratio $J/M^2$, where $J$ is the angular
   momentum, and the degree of differential rotation $\hat{A}$, where
   for $\hat{A}\rightarrow\infty$ uniform rotation is recovered. All
   the initial models have been computed with a polytropic EOS with
   $K=100$ and $N=1$. The last column shows instead the ratio
   $J_{BH}/M_{BH}^2$ for the BH formed after the collapse of the
   Sub-Kerr models computed using Eq. (5.2) in
   Ref.~\cite{whisky}.}
 \begin{ruledtabular}
 \begin{tabular}{ccccccccccc}
Model & $\rho_c \times 10^{-3}$ & $\rho_{\mathrm{max}} \times 10^{-3}$ & $r_p/r_e$ & $M/M_{\odot}$ & $R_e/M$& $\Omega_c$ & $T/|W|$ & $J/M^2$ & $\hat{A}$ & $J_{BH}/M_{BH}^2$\\
\hline
$\mathtt{A1}$ & 3.0623 & 6.8920 & 0.23 & 1.7626 & 3.5424  & 0.51891 & 0.18989 & 0.75004 & 0.6 & 0.74\\
$\mathtt{A2}$ & 3.0623 & 4.0236 & 0.33 & 2.2280 & 3.5316  & 0.21752 & 0.21705 & 0.81507 & 1.0 & 0.81\\
$\mathtt{A3}$ & 3.0623 & 3.0623 & 0.33 & 2.6127 & 4.1111  & 0.10859 & 0.23163 & 0.88474 & 1.4 & 0.88\\
$\mathtt{B1}$ & 0.4630 & 0.4632 & 0.39 & 1.9009 & 8.8185  & 0.03723 & 0.21509 & 1.08650 & 1.0 & 
 \end{tabular}
 \end{ruledtabular}
 \end{table*}

\begin{figure}
  \begin{center}
  \includegraphics[width=0.45\textwidth]{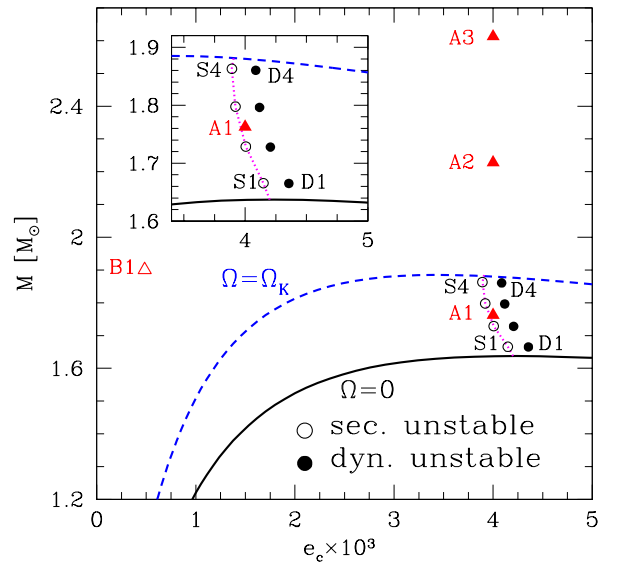}
  \end{center}
  \caption{\label{fig:diffrot_idata}Comparison between our initial
    models (filled triangles; see Table~\ref{tab:diffrot_id}) and the
    uniformly rotating models studied in Ref.~\cite{whisky}. Here we
    plot the gravitational mass $M$ as a function of the central
    energy density $e_c$. The solid, dashed, and dotted lines
    correspond to the sequence of nonrotating models, the sequence of
    models rotating at the mass shedding limit, and the sequence of
    uniformly rotating models that are at the onset of the secular
    instability to axisymmetric perturbations. Also shown are the
    secularly (open circles) and dynamically unstable (filled circles)
    initial models used in Ref.~\cite{whisky} (see inset).}
\end{figure}

\begin{figure*}
  \begin{center}
  \includegraphics[width=0.35\textwidth]{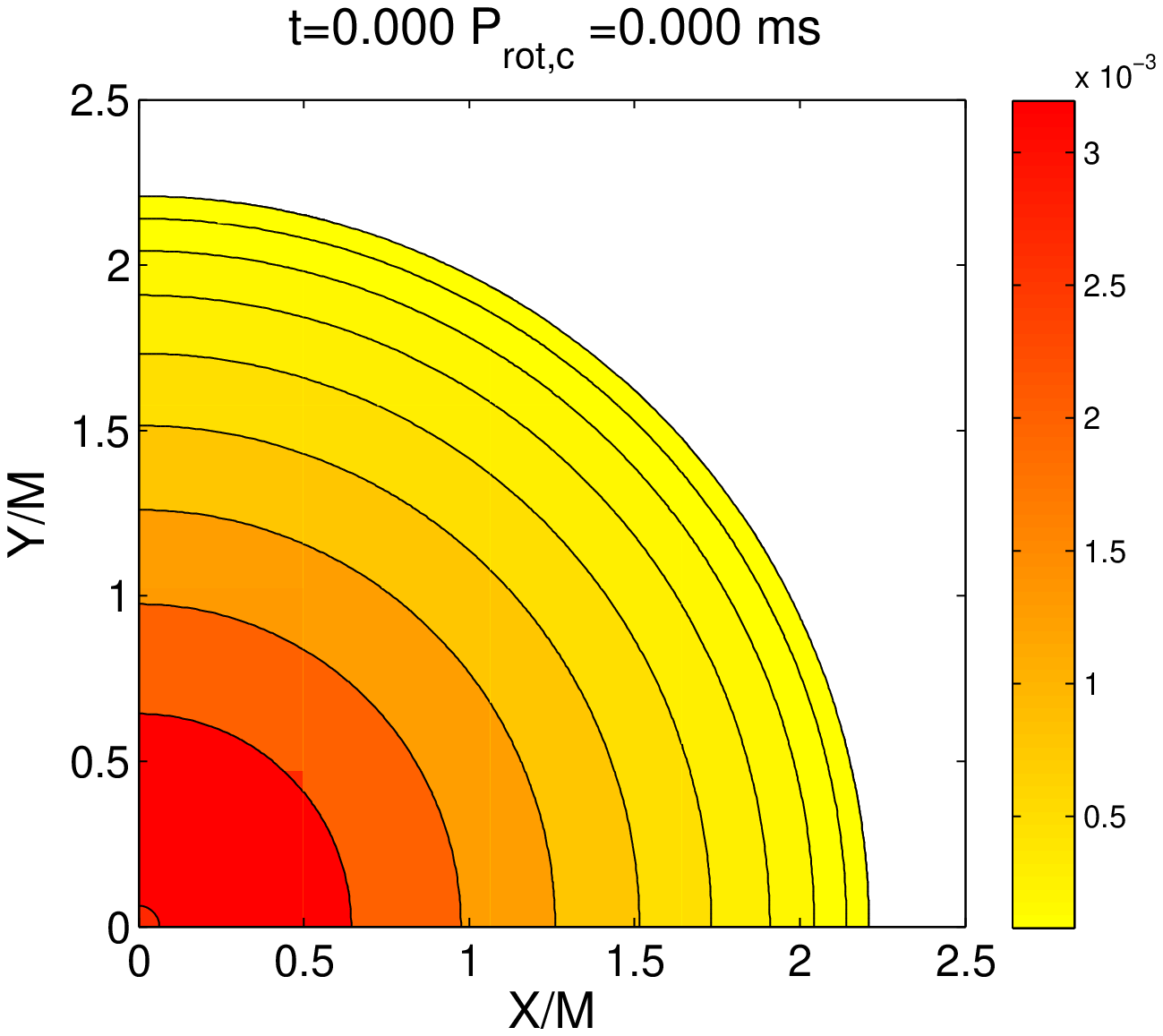}
  \hskip 1.0cm
  \includegraphics[width=0.35\textwidth]{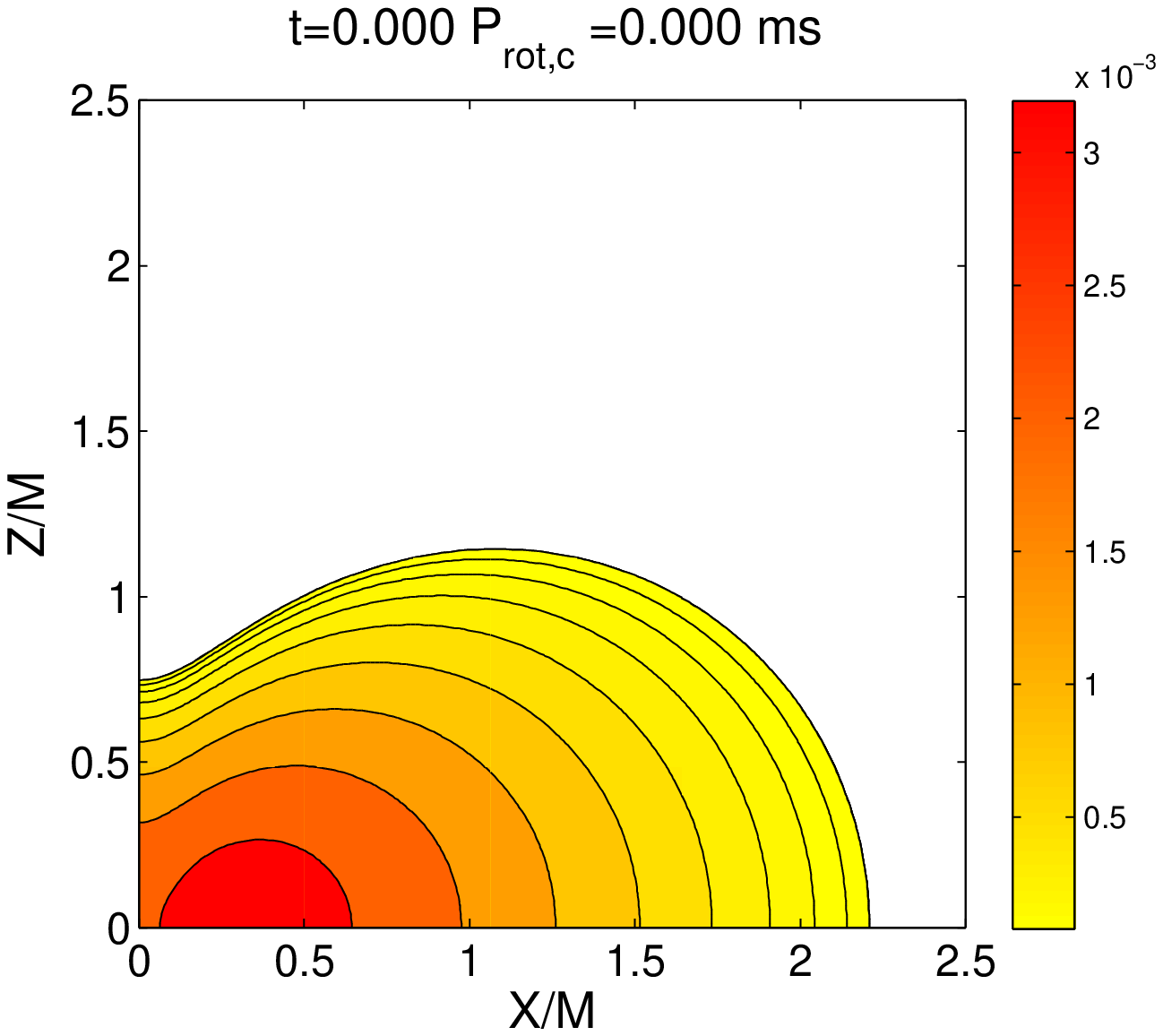}
  \vskip 0.1cm
  \includegraphics[width=0.35\textwidth]{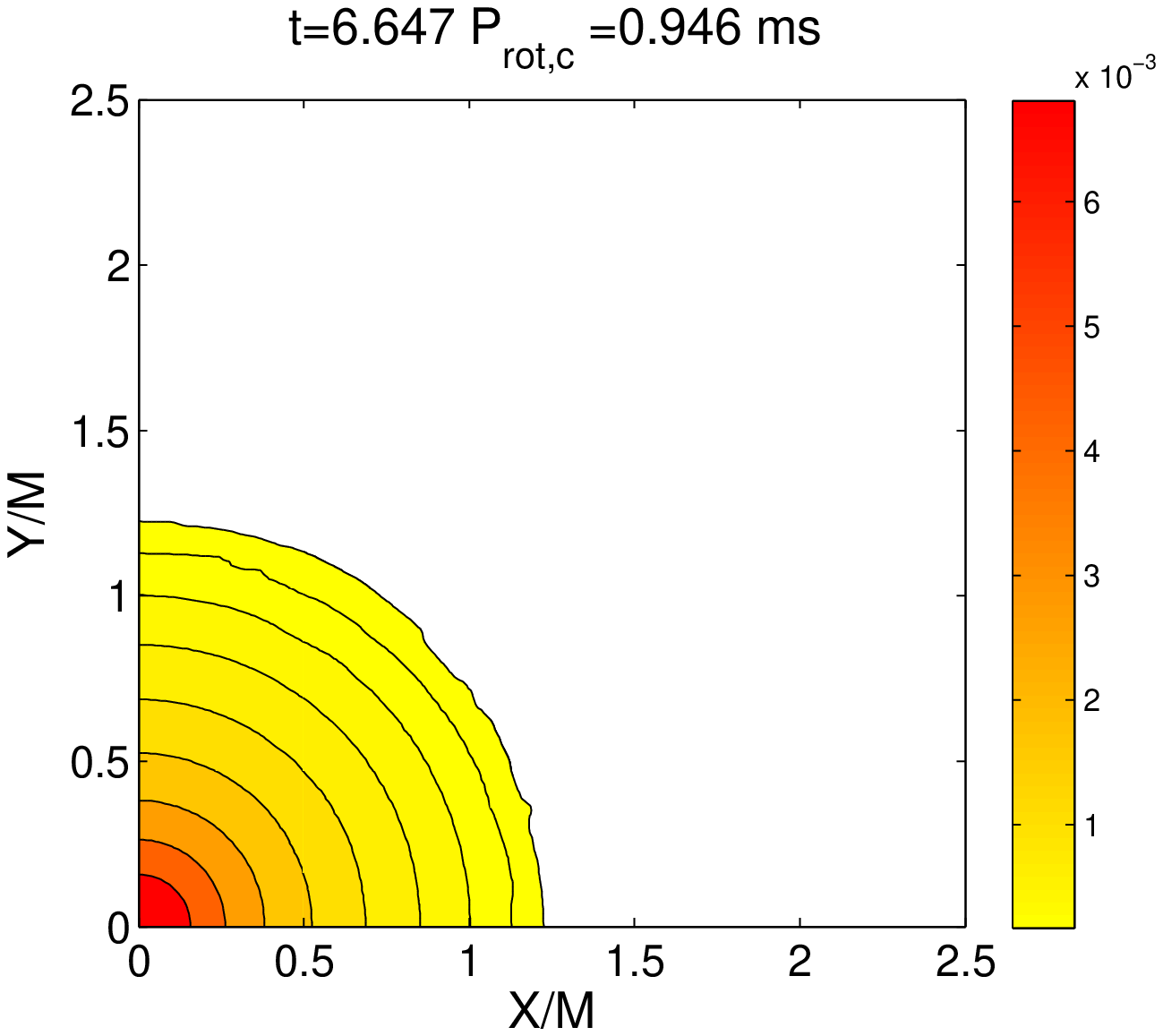}
  \hskip 1.0cm
  \includegraphics[width=0.35\textwidth]{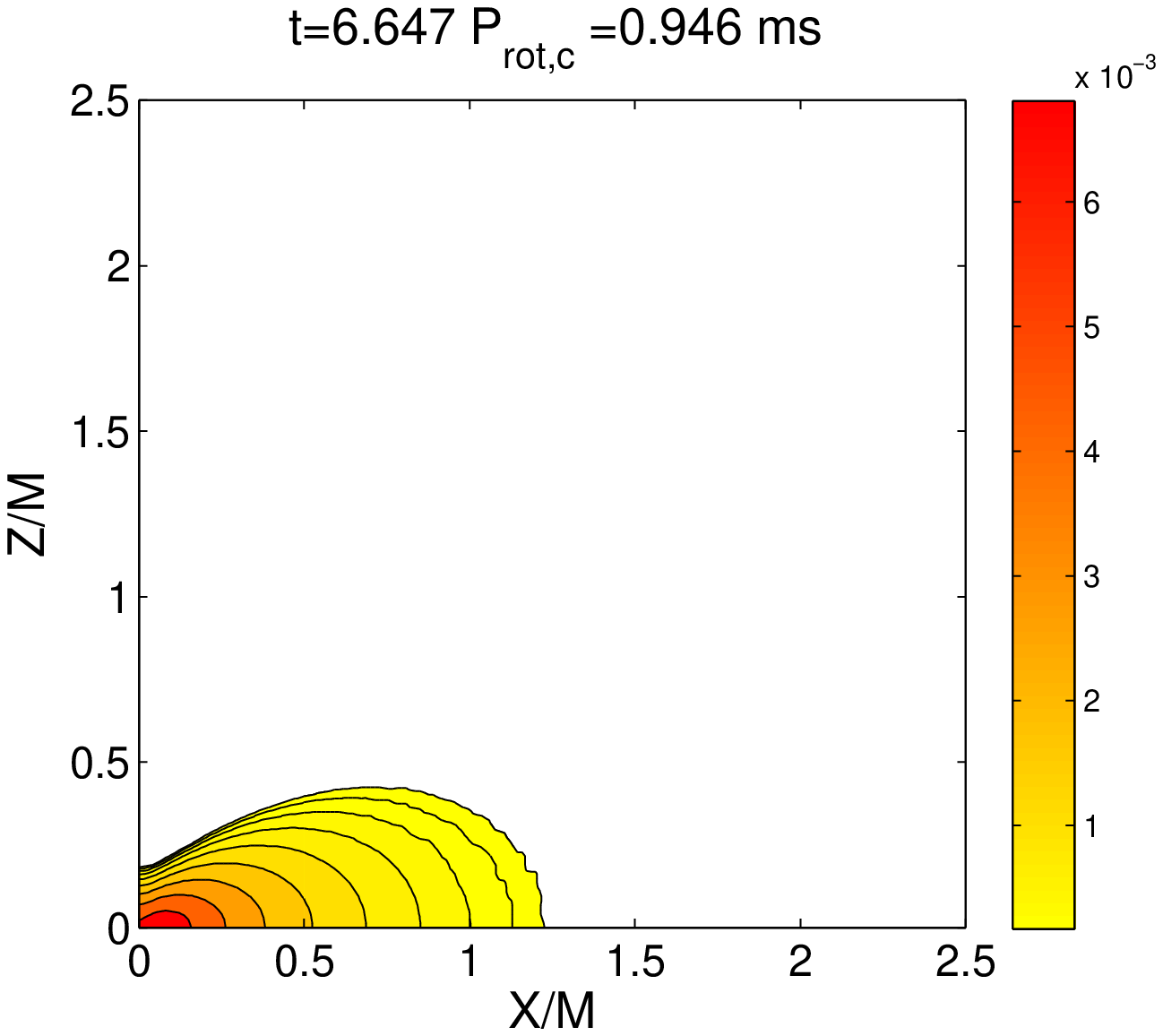}
  \vskip 0.1cm
  \includegraphics[width=0.35\textwidth]{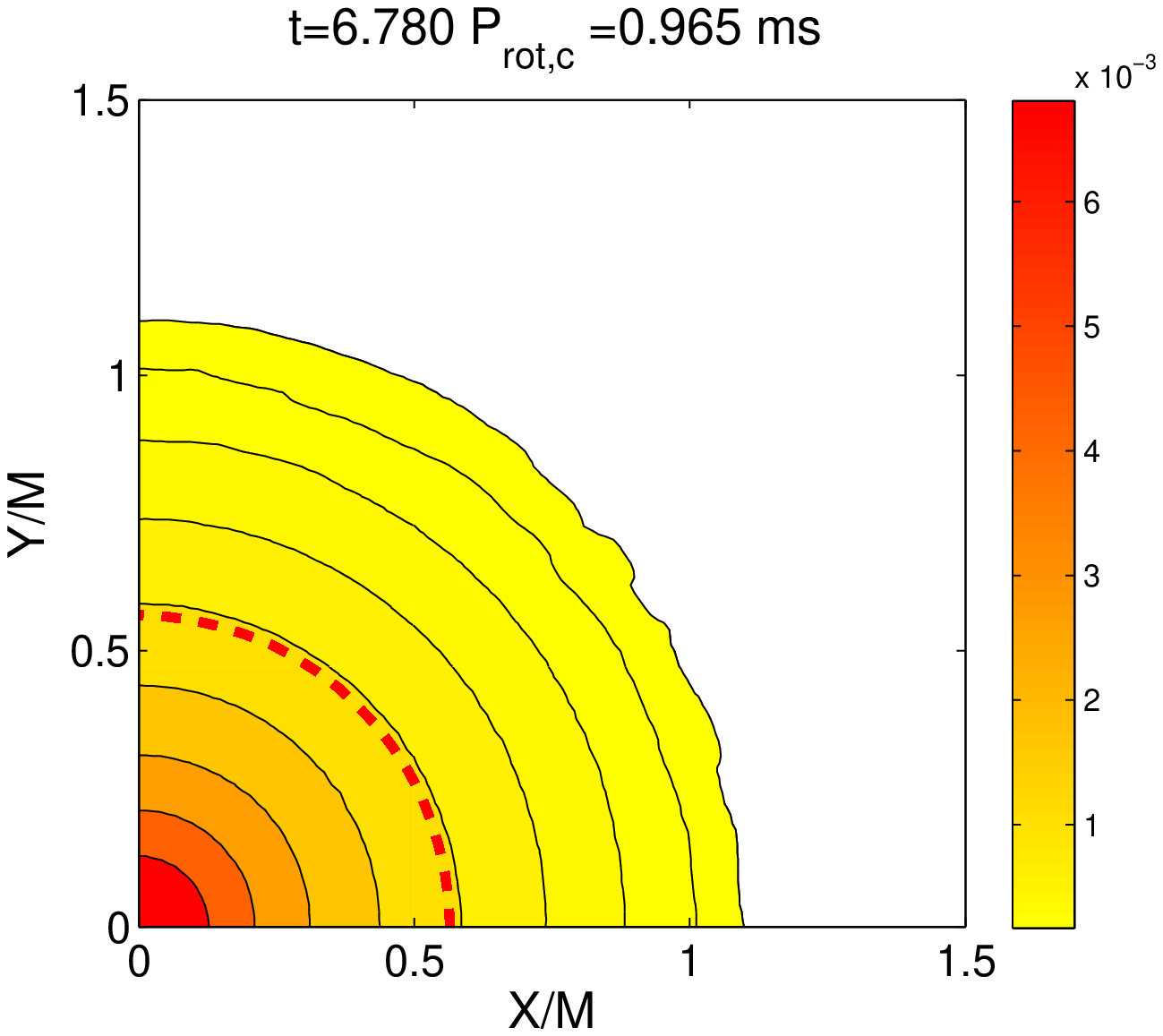}
  \hskip 1.0cm
  \includegraphics[width=0.35\textwidth]{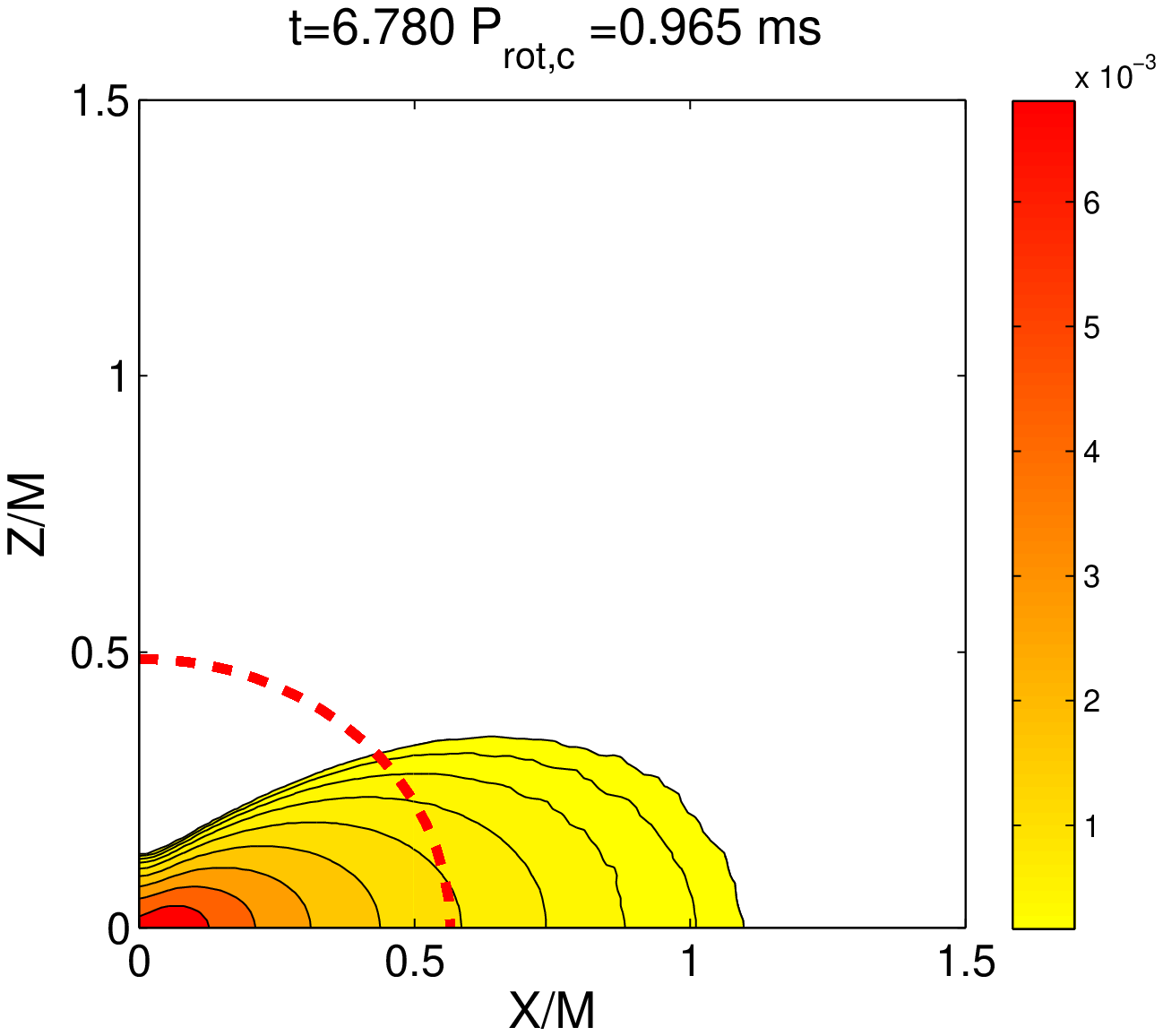}
  \vskip 0.1cm
  \includegraphics[width=0.35\textwidth]{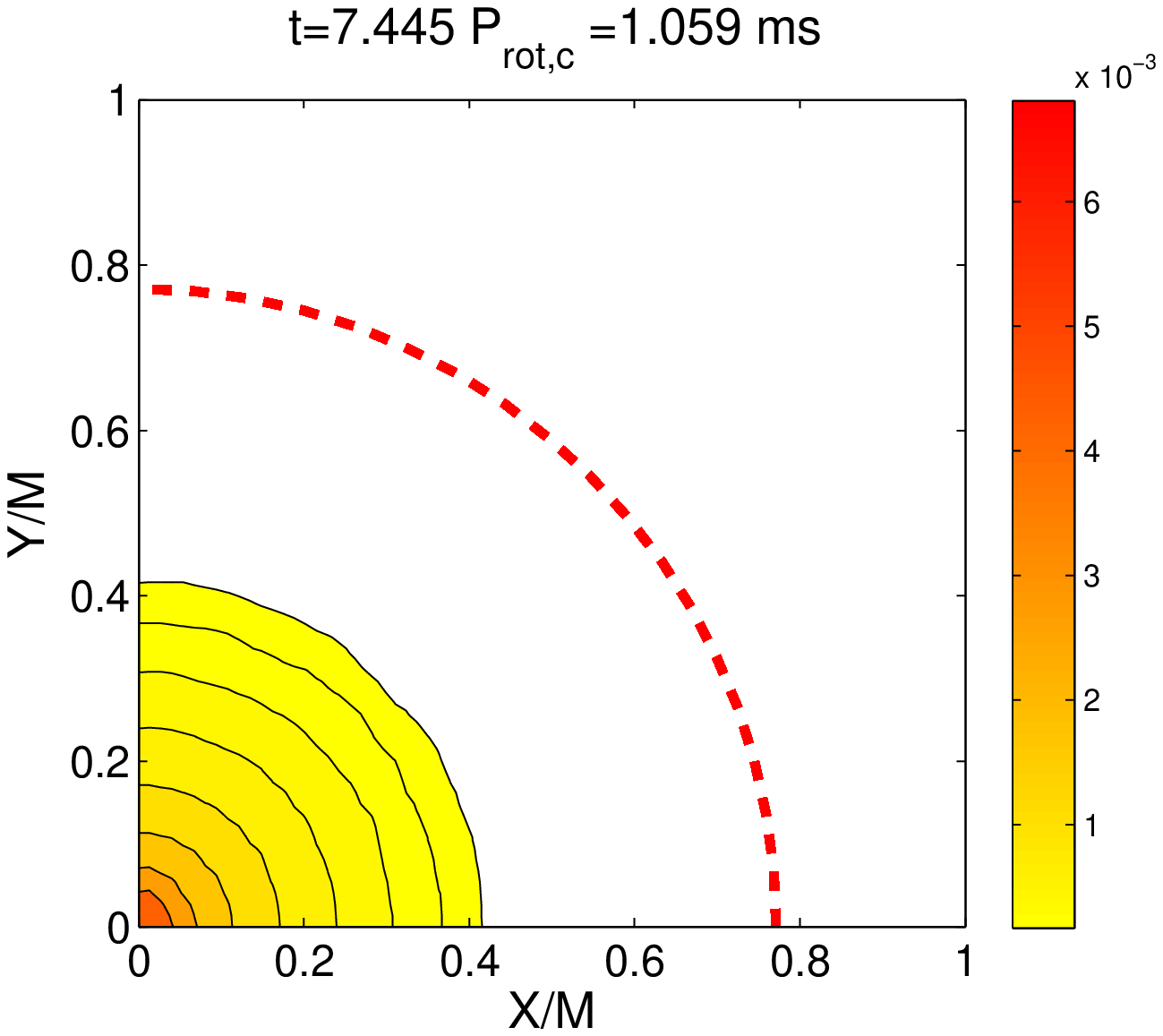}
  \hskip 1.0cm
  \includegraphics[width=0.35\textwidth]{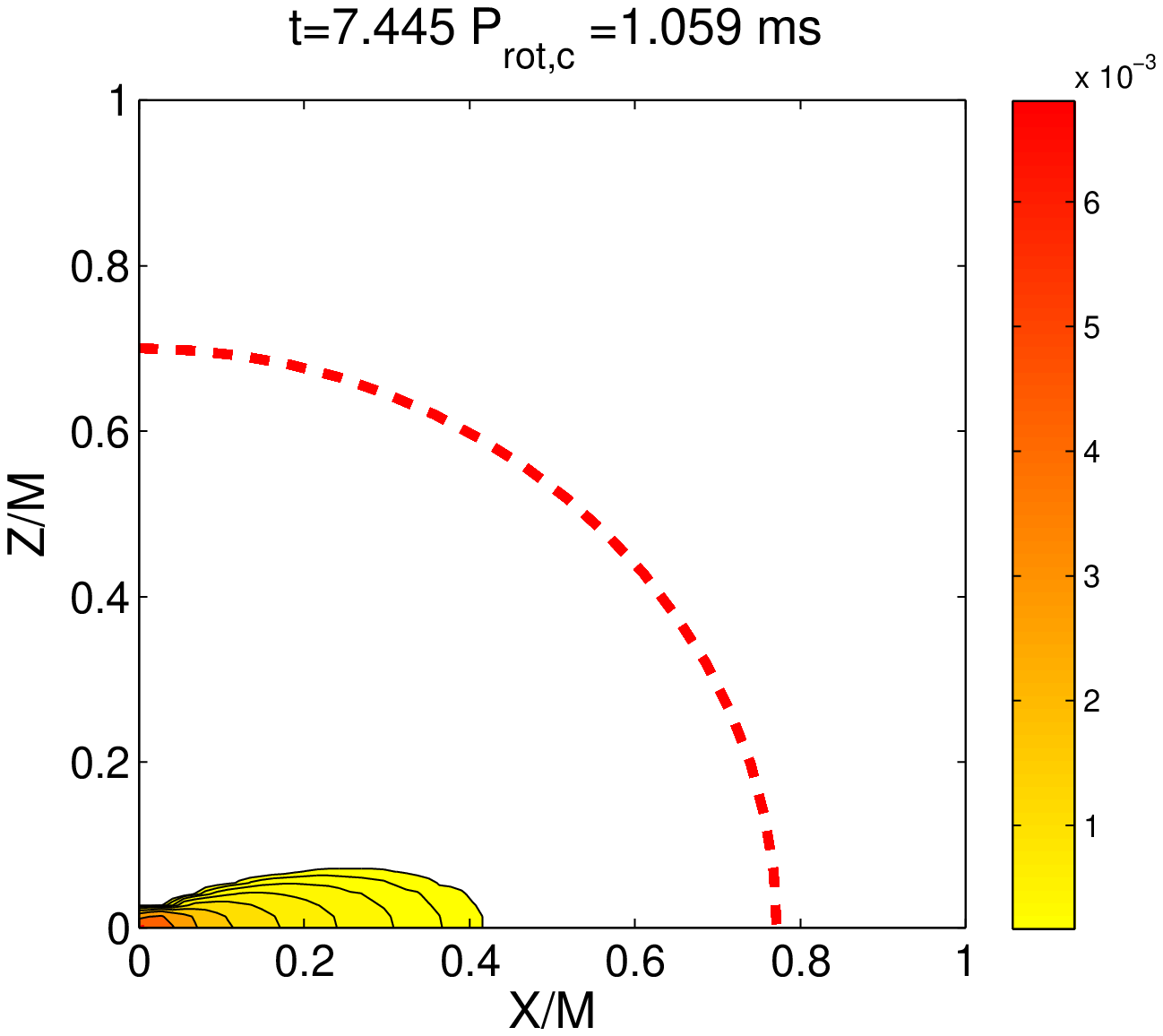}
  \end{center}
  \caption{\label{fig:subkerr_dynamics_part1}Snapshots of the
    rest-mass density $\rho$ in the equatorial plane (left panels) and
    in the $xz$ plane (right panels) for model {$\mathtt{A2}$}. The contour
    lines are drawn for $\rho=10^{-(0.2 j+0.1)}\rho_{\text{max}}$
    for $j=0,1,\ldots,8$, where $\rho_{\text{max}}$ is the maximum of
    $\rho$. After the formation of the AH (dashed line) at time
    $t_{\rm ah}=6.71P_{\text{rot,c}}$, $\rho_{\text{max}}$ is the maximum of
    $\rho$ at $t=t_{\rm ah}$. Time is normalized to the initial central
    rotation period of the star, $P_{\text{rot,c}}=13M$.}
\end{figure*}

%%%%%%%%%%%%%%%%%%%%%%%%%%%%%%%%%%%%%%%%%%%%%%%%%%%%%%%%%%%
\subsection{Supra- and sub-Kerr initial data}
%%%%%%%%%%%%%%%%%%%%%%%%%%%%%%%%%%%%%%%%%%%%%%%%%%%%%%%%%%%

We have investigated the dynamics of differentially rotating
collapsing compact stars by focusing on three sub-Kerr, dynamically
unstable models and on one supra-Kerr, dynamically stable model. For
the latter, the collapse was triggered through an artificially large
pressure depletion of $99\%$, as was done in Ref.~\cite{duez04}. All
models are constructed with the polytropic EOS with $N=1$ and $K=100$.
The three dynamically unstable models are labeled $\mathtt{A1}$ to
$\mathtt{A3}$ and are shown as filled triangles in
Fig.~\ref{fig:diffrot_idata} while their detailed properties are
displayed in Table~\ref{tab:diffrot_id}. The central rest-mass density
of the three models is chosen to be the same as the central rest-mass
density of the maximum mass nonrotating model for this EOS. The degree
of differential rotation varies from $\hat{A}=0.6$ to
$\hat{A}=1.4$. The maximum rest-mass density increases with respect to
the central density, as differential rotation becomes stronger
(\textit{i.e.}, as the relative length scale $\hat{A}$ becomes
smaller), as can be seen by comparing the values in the second and
third columns of Table~\ref{tab:diffrot_id}. All three models have
comparable values of $J/M^2$ ($0.75$ to $0.88$), of the ratio of
rotational kinetic energy to gravitational binding energy $T/|W|$
($0.19$ to $0.23$) and of the total (gravitational) mass $M$ ($1.8$ to
$2.6$), while they differ significantly in radius ($6.4$ to $11$) and
central angular velocity ($0.52$ to $0.11$).  Even though the
axisymmetric stability of these models cannot be determined from a
turning-point method, the numerical simulations performed and
discussed later on, showed that these models are indeed dynamically
unstable and collapse without the need of a pressure depletion.

Finally, the fourth model studied, model $\mathtt{B1}$ in
Table~\ref{tab:diffrot_id}, is shown as a empty triangle in
Fig.~\ref{fig:diffrot_idata} and represents a stable supra-Kerr model,
with comparable mass and $T/|W|$-ratio as models $\mathtt{A1}$ to
$\mathtt{A3}$, but with much smaller central rest-mass density and
$J/M^2=1.09$. The complete set of initial data is displayed in
Fig.~\ref{fig:diffrot_idata}, which reports the total mass versus the
energy density at the center of the star $e_c$. Also shown as useful
references are the sequence of nonrotating models (solid line), the
sequence of models rotating at the mass-shedding limit (dashed line)
and the sequence of models that are at the onset of the secular
instability to axisymmetric perturbations (dotted line). Similarly,
indicated respectively with open and filled circles, are the secularly
($\mathtt{S1-S4}$) and dynamically unstable ($\mathtt{D1-D4}$)
uniformly rotating models used in the collapse simulations discussed
in Ref.~\cite{whisky} (see Fig.~\ref{fig:diffrot_idata} inset).

%%%%%%%%%%%%%%%%%%%%%%%%%%%%%%%%%%%%%%%%%%%%%%%%%%%%%%%
\section{\label{sec:dynamics}Dynamics of the collapse}
%%%%%%%%%%%%%%%%%%%%%%%%%%%%%%%%%%%%%%%%%%%%%%%%%%%%%%%

We next discuss the dynamics of the matter during the collapse of the
initial stellar models described in the preceding section. All the
models were studied with different resolutions and using fixed mesh
refinement techniques. In the case of the sub-Kerr models
$\mathtt{A1}$, $\mathtt{A2}$, and $\mathtt{A3}$, up to seven refinement
levels were used, in order to be able to extract the
gravitational-wave signal in a region sufficiently distant from the
sources.

The supra-Kerr model $\mathtt{B1}$, on the other hand, was studied
using three levels because the dynamics is not limited to a small
central region of the computational domain (the process follows
several bounces and subsequent collapses) and so the finest refinement
level was set to be larger with respect to the one used for the
sub-Kerr models in order to reduce the numerical error. The outer
boundaries were then moved at those distances that were
computationally affordable, but not sufficiently far away to allow for
gravitational-wave extraction as in the sub-Kerr cases. For this
model, in fact, we resort to a gravitational-wave extraction via the
quadrupole formula, as discussed in Sec.~\ref{ssec:suprakerr_waves}.

An ideal-fluid EOS, $P=\rho \epsilon (\Gamma-1)$, with $\Gamma=2$ was
used during the evolution of all the models.

\begin{figure}
  \begin{center}
    \includegraphics[width=0.45\textwidth]{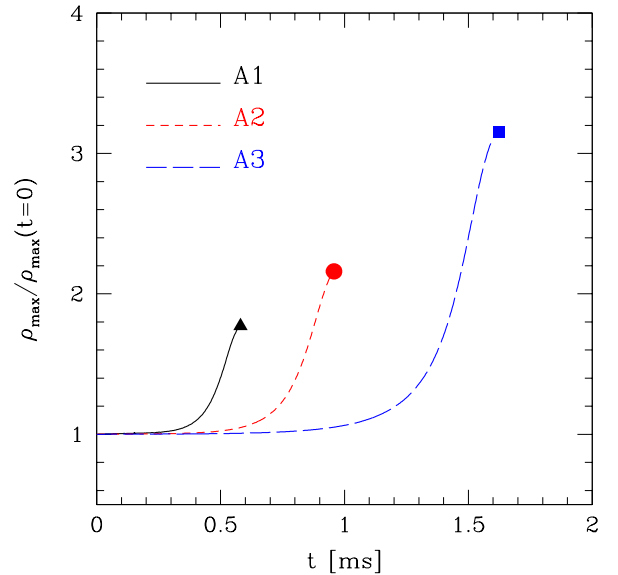}
  \end{center}
  \caption{\label{fig:subkerr_rhomax}Maximum of the rest-mass density
    $\rho$ normalized to its initial value. The triangle, the circle
    and the square represent the moment at which an apparent horizon
    is formed for models {$\mathtt{A1}$}, {$\mathtt{A2}$}, and
    {$\mathtt{A3}$}, respectively. This figure should be compared with
    the corresponding Fig.~\ref{fig:suprakerr_rhomax}, which refers to
    the supra-Kerr model $\mathtt{B1}$.}
\end{figure}

\subsection{\label{ssec:subkerr}Sub-Kerr Collapse}

All the three sub-Kerr models considered $\mathtt{A1}$, $\mathtt{A2}$,
$\mathtt{A3}$ show the same qualitative dynamics, with the
gravitational collapse leading to a central black hole. All of them
were evolved in equatorial and $\pi/2$ symmetry (\textit{i.e.}, we
considered the region $\{x>0,y>0,z>0\}$ applying reflection symmetry
at $z=0$ and a rotational symmetry at $x=0$ and $y=0$) since they did
not show the development of nonaxisymmetric instabilities (when
evolved without symmetries at lower resolutions), in a way similar to
the uniformly rotating models studied in Ref.~\cite{whisky}.

\begin{figure}
  \begin{center}
  \includegraphics[width=0.45\textwidth]{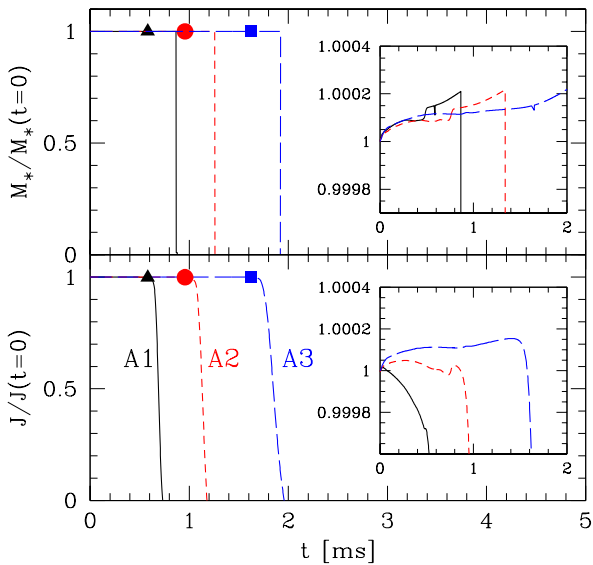}
  \end{center}
  \caption{\label{fig:subkerr_restmass}Total rest-mass $M_*$ and
    angular momentum $J$ {of the matter}, normalized at $t=0$ for the
    three different models {$\mathtt{A1}$} (solid line),
    {$\mathtt{A2}$} (short-dashed line), and {$\mathtt{A3}$}
    (long-dashed line). The triangle, the circle, and the square
    represent the moment at which an apparent horizon is formed for,
    respectively, model {$\mathtt{A1}$}, {$\mathtt{A2}$}, and
    {$\mathtt{A3}$}.
  }
\end{figure}

Because of the similar behavior, we concentrate here only on the
description of model {$\mathtt{A2}$}. The results shown here were
produced on a grid with boundaries located at
$[0,137.9M]\times[0,137.9M]\times[0,137.9M]$ with a resolution ranging
from $\Delta x^i=0.86M$ on the coarsest grid to $\Delta x^i=0.013M$ on
the finest one, using seven refinement levels. The boundaries of the
finest grid were chosen in order to include all the star and thus to
reduce numerical errors. The results shown here were obtained without
the introduction of any initial perturbation, except from the
truncation error, but a similar dynamics was obtained when the
collapse was triggered by reducing the pressure by~$2\%$, as was done
in the case of uniformly rotating models in Ref.~\cite{whisky}.

\begin{figure*}
  \begin{center}
  \includegraphics[width=0.33\textwidth]{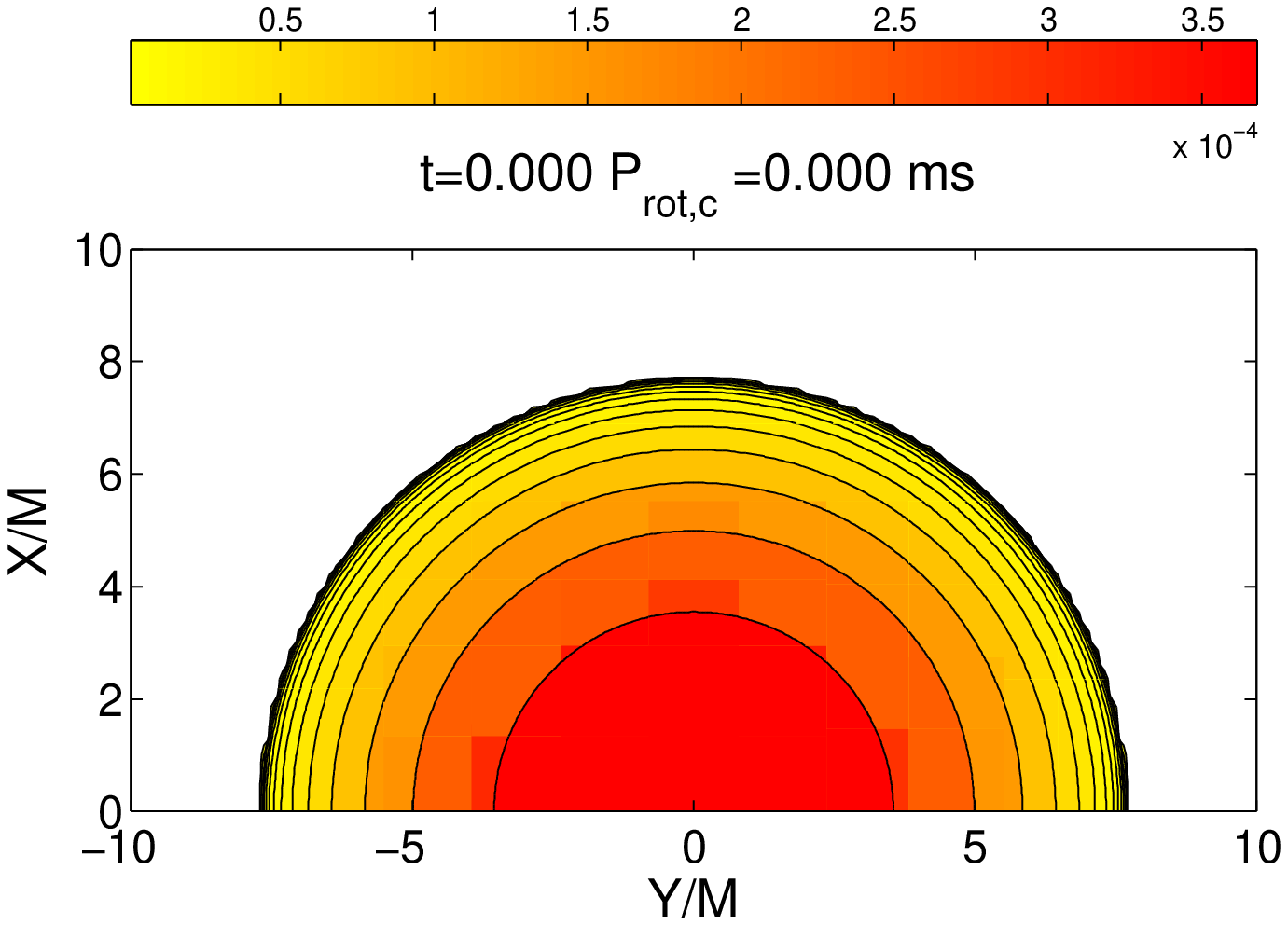}
  \hskip 1.0cm
  \includegraphics[width=0.33\textwidth]{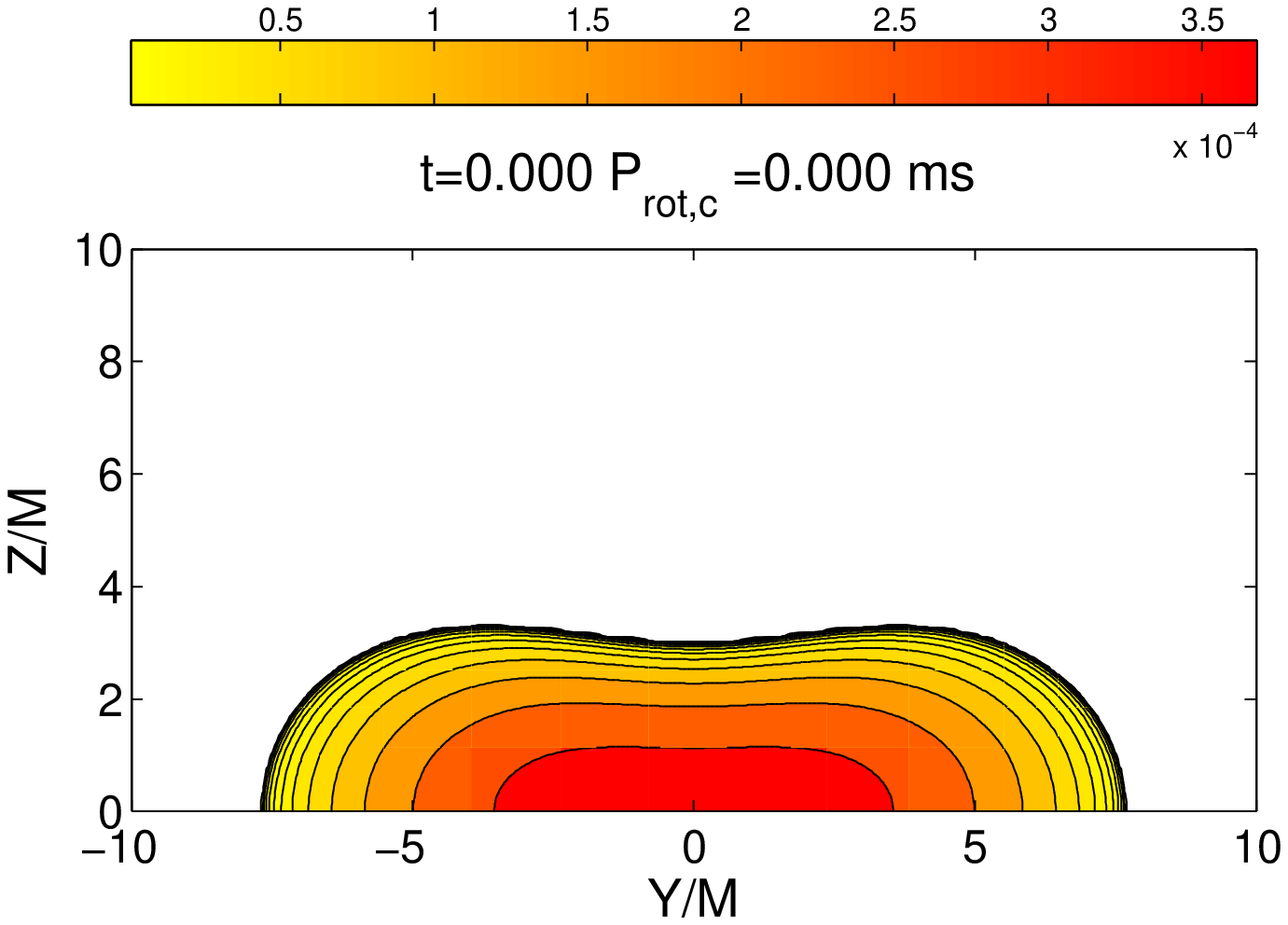}
  \vskip 0.25cm
  \includegraphics[width=0.33\textwidth]{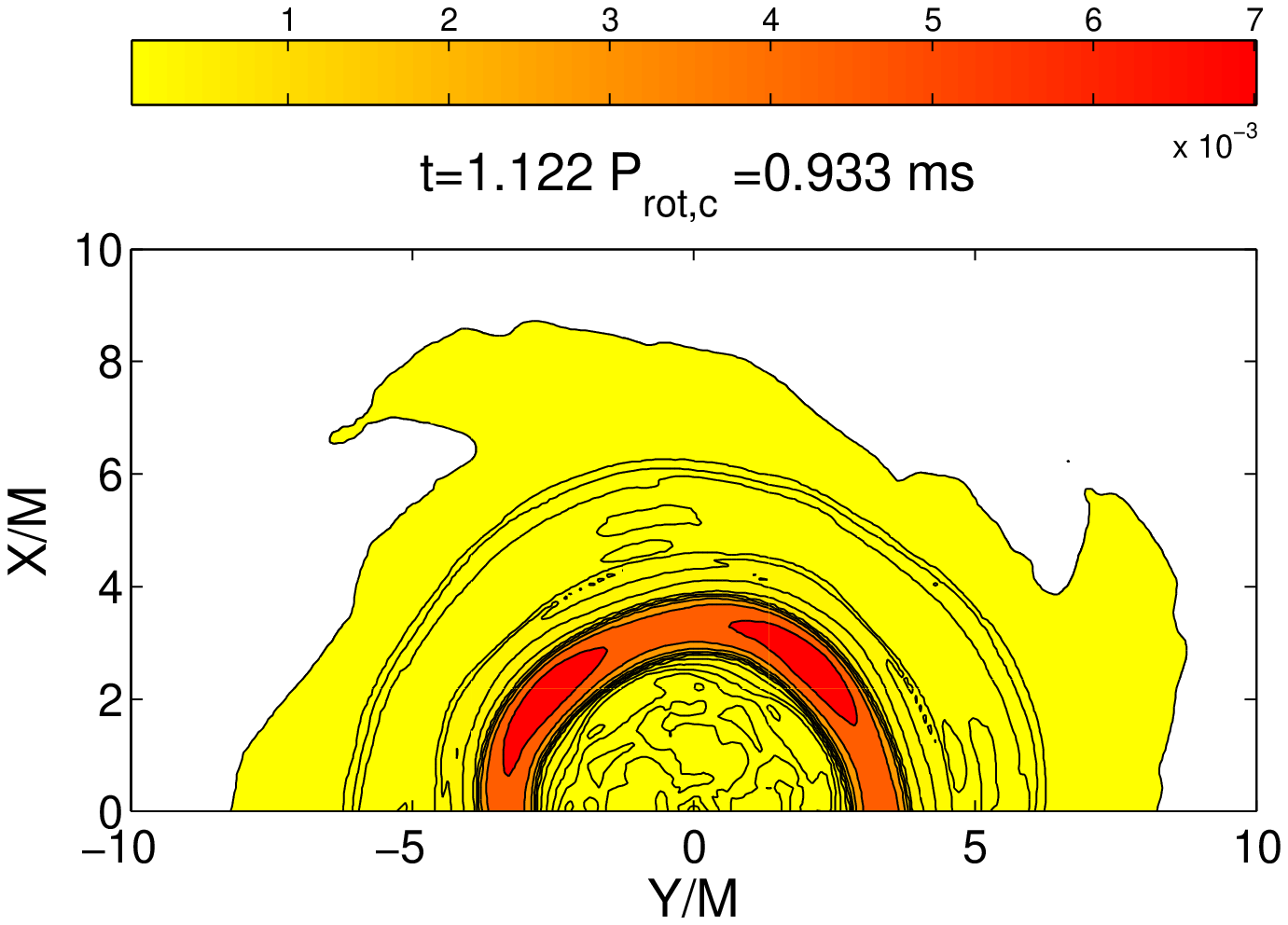}
  \hskip 1.0cm
  \includegraphics[width=0.33\textwidth]{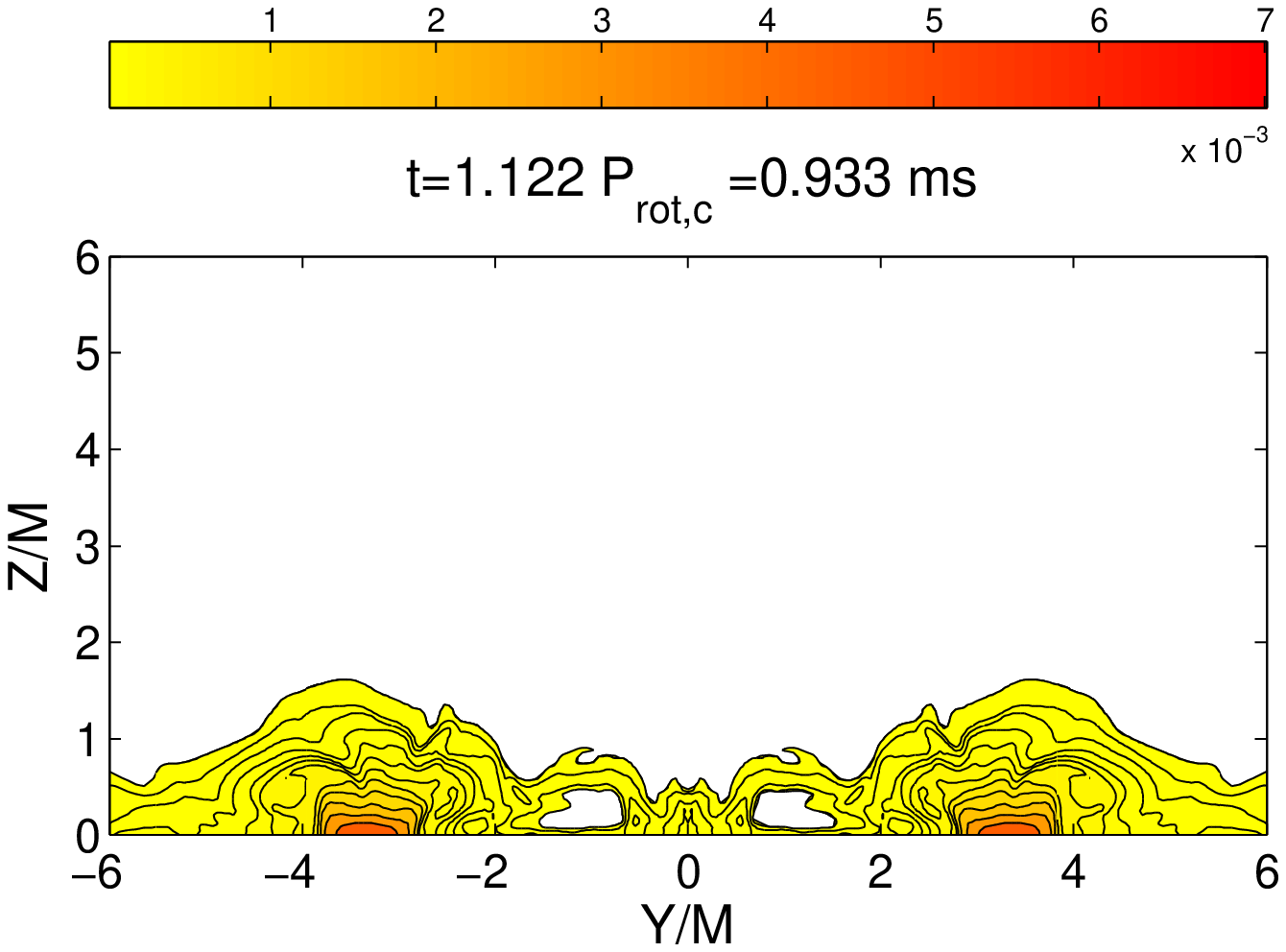}
  \vskip 0.25cm
  \includegraphics[width=0.33\textwidth]{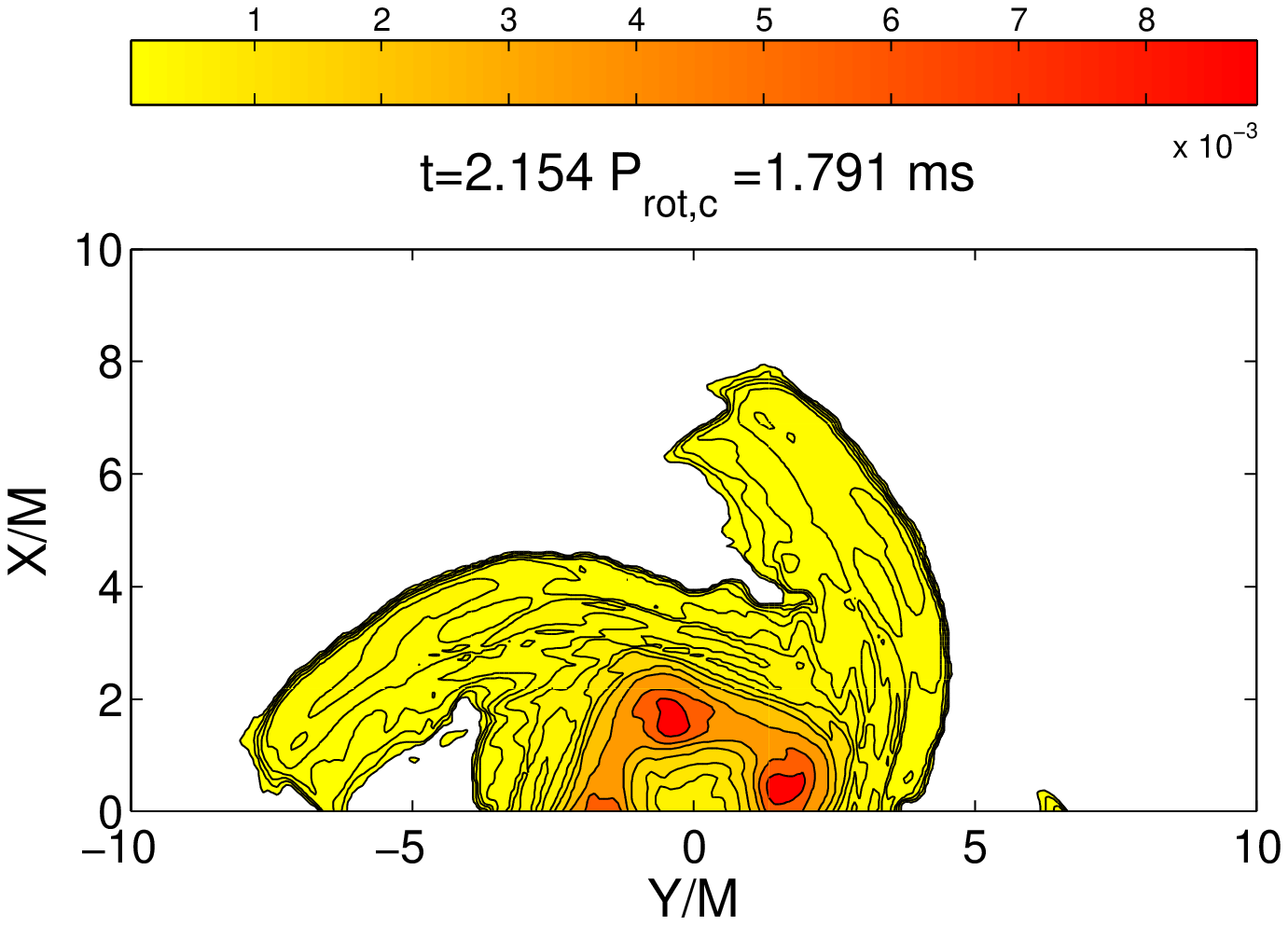}
  \hskip 1.0cm
  \includegraphics[width=0.33\textwidth]{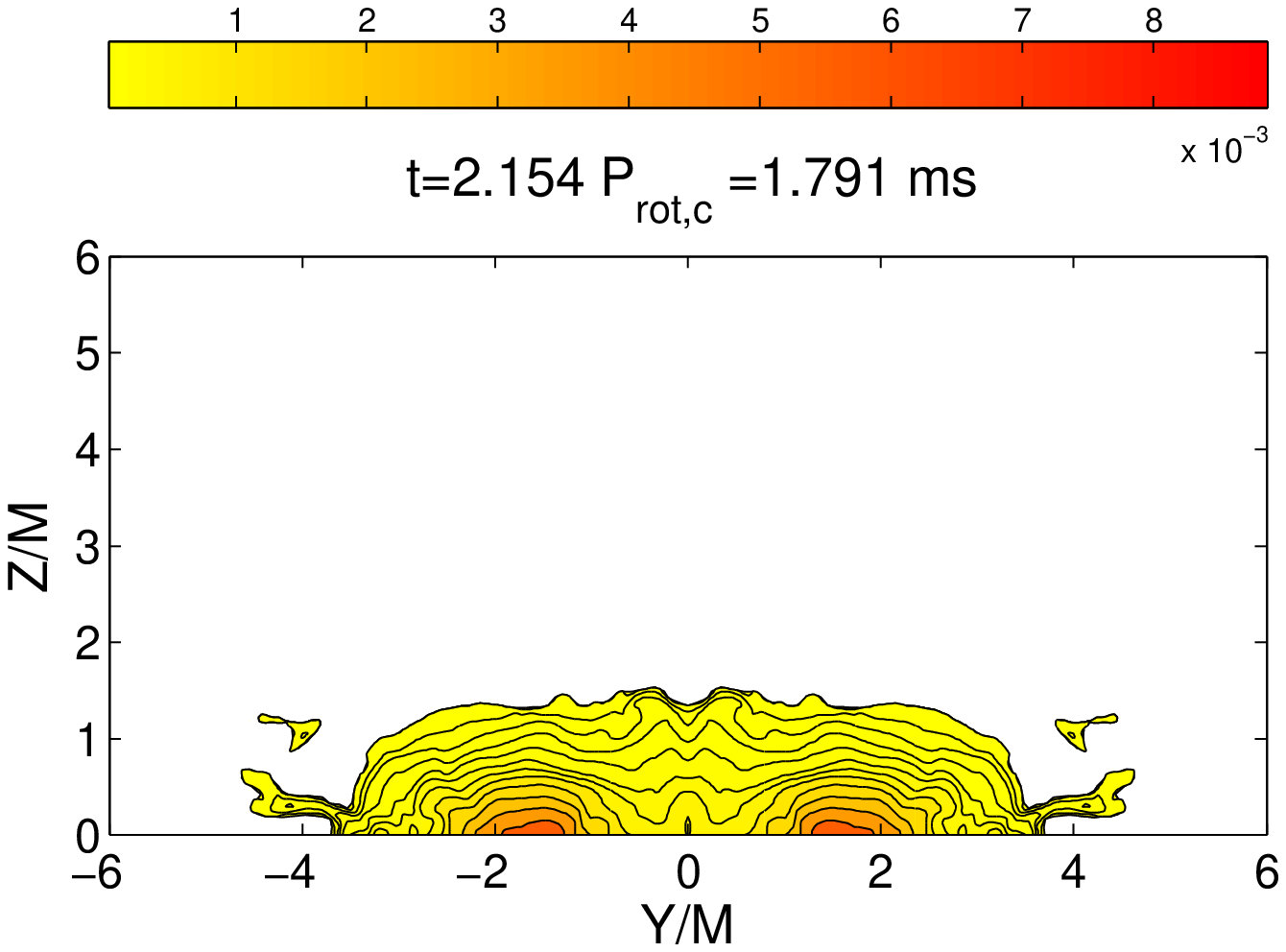}
  \vskip 0.25cm
  \includegraphics[width=0.33\textwidth]{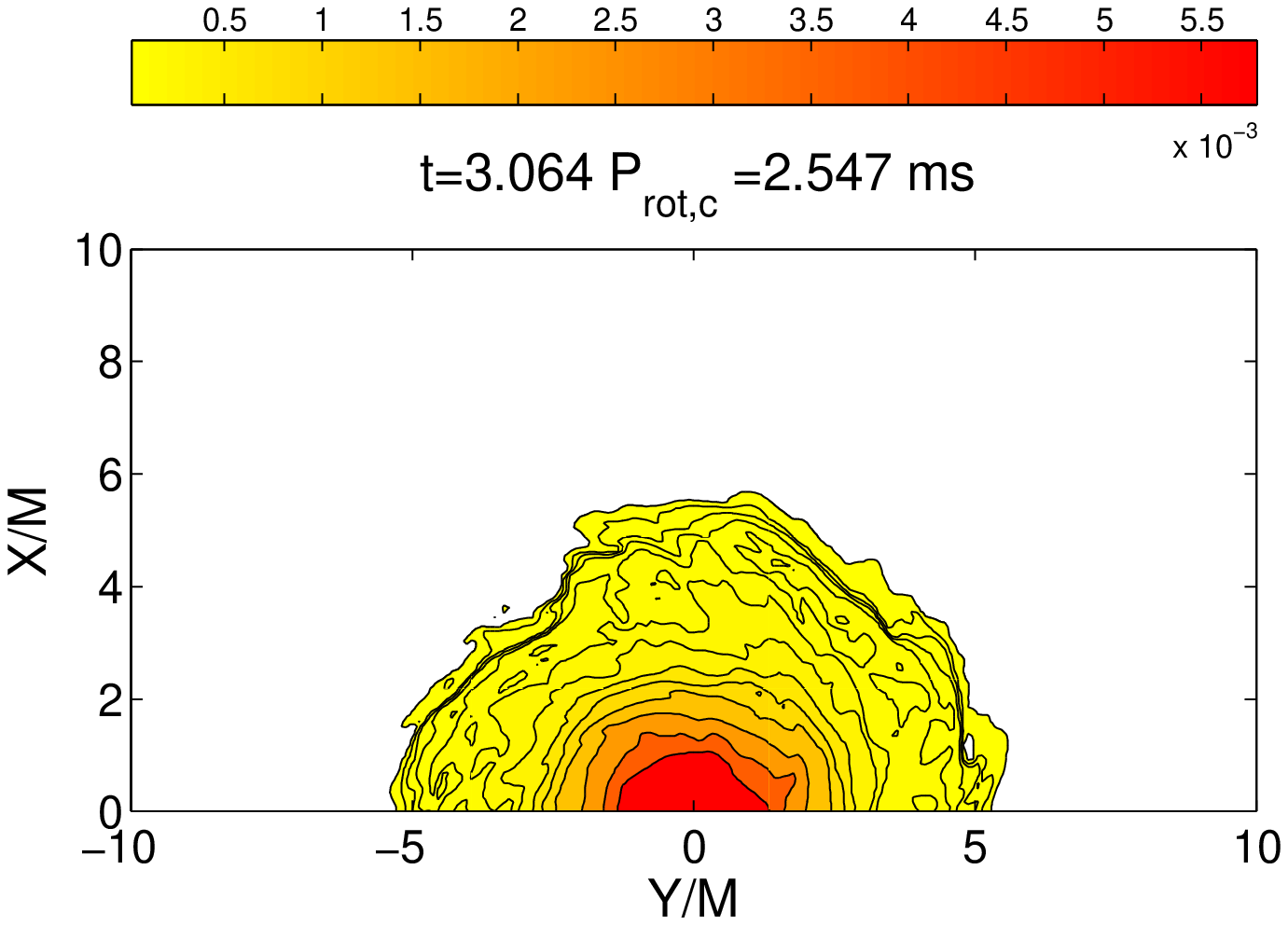}
  \hskip 1.0cm
  \includegraphics[width=0.33\textwidth]{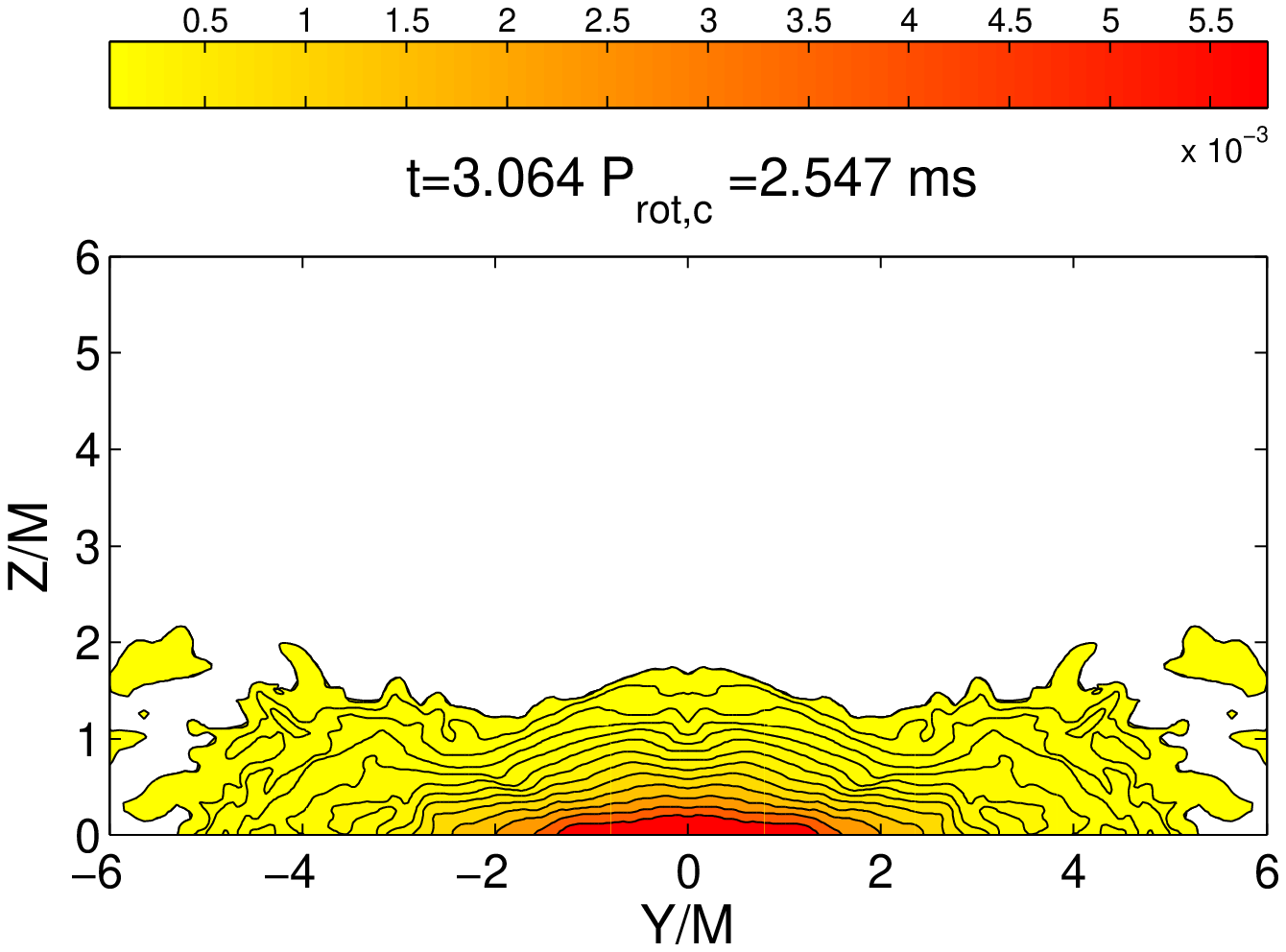}
  \vskip 0.25cm
  \includegraphics[width=0.33\textwidth]{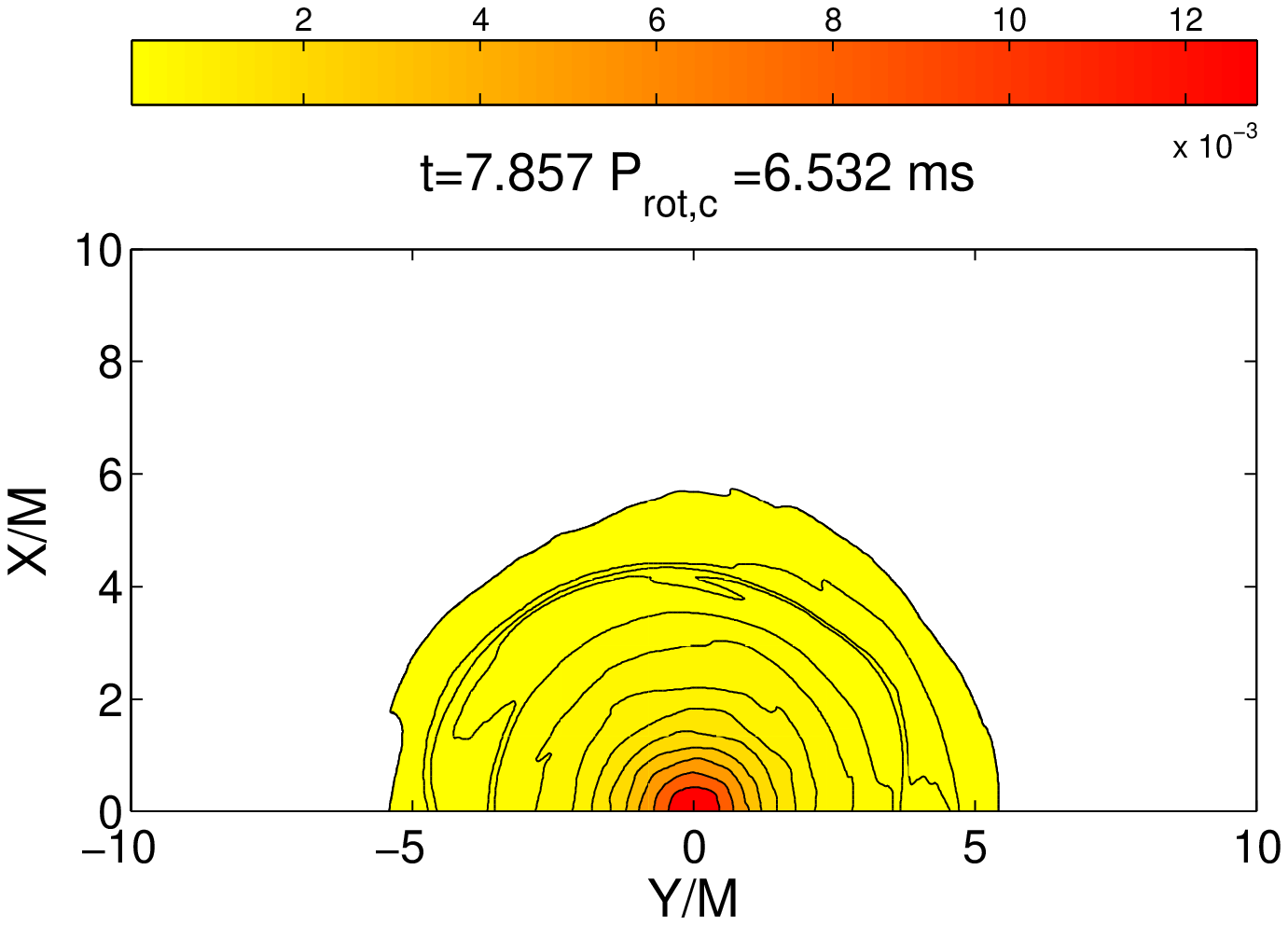}
  \hskip 1.0cm
  \includegraphics[width=0.33\textwidth]{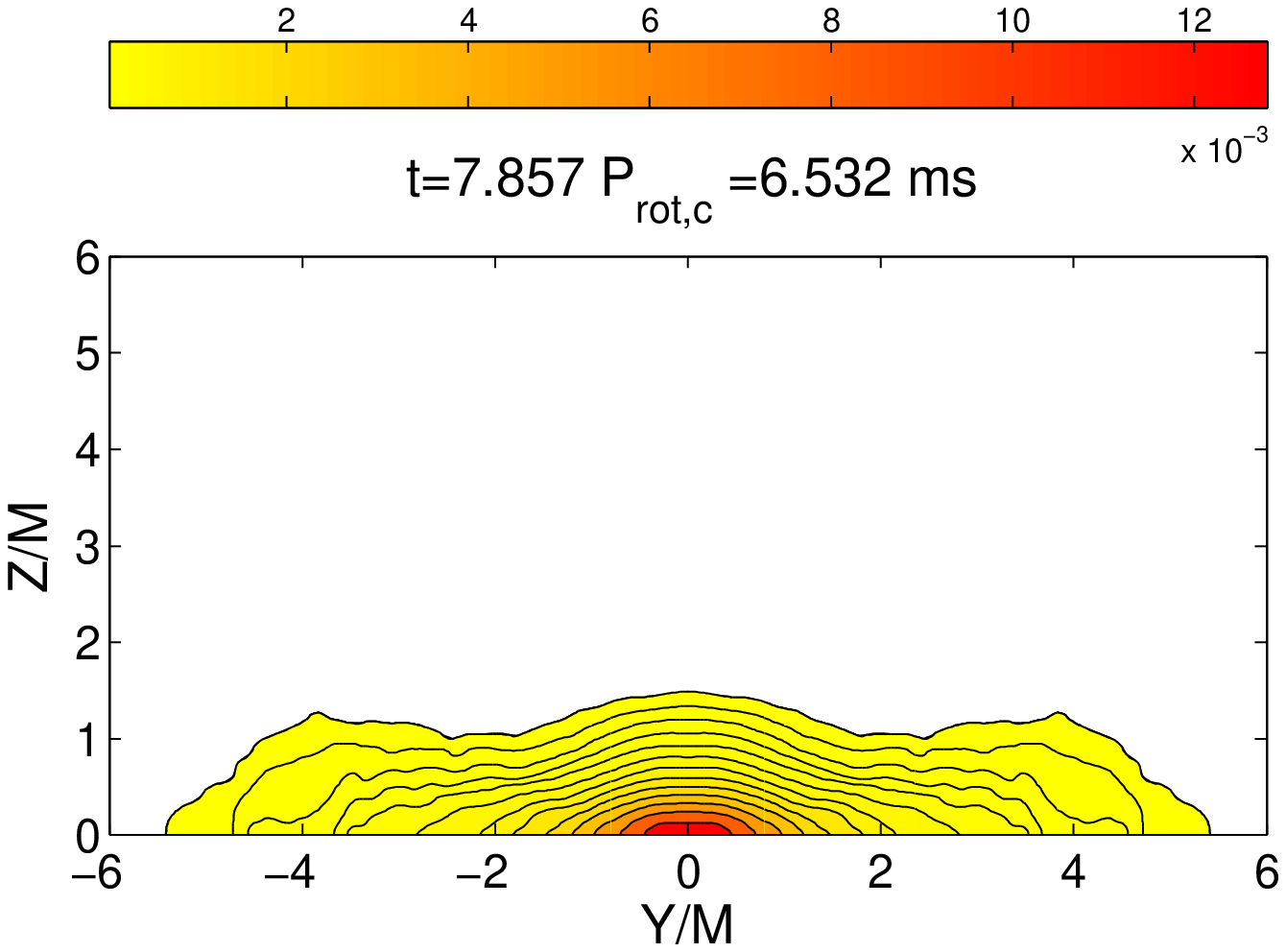}
  \end{center}
  \caption{\label{fig:suprakerr_dynamics}Snapshots of the rest-mass
    density $\rho$ in the equatorial $xy$ plane (left panels) and $yz$
    plane (right panels) for model $\mathtt{B1}$. The contour lines
    are drawn for $\rho=10^{-(0.2 j+0.1)}\mathrm{max}(\rho)$ for
    $j=0,1,\ldots,12$. Time is normalized to the initial central
    rotational period of the star, $P_{\text{rot,c}}=89M$.}
\end{figure*}

As one can see in the first row of panels in
Fig.~\ref{fig:subkerr_dynamics_part1}, where we show the isodensity
contours in the $(x,y)$ (\ie equatorial) and $(x,z)$ planes, the star
has a toroidal shape due to its strong differential rotation. Its
evolution is rather similar to what was already observed for the
uniformly rotating models and especially for model $\mathtt{D4}$ in
Ref.~\cite{whisky}. The collapse is axisymmetric and leads to the
formation of a black hole, as indicated by the appearance of an
apparent horizon (AH)~\cite{Thornburg2003:AH-finding}. The AH,
represented by a dashed line in Fig.~\ref{fig:subkerr_dynamics_part1},
is found at $t=6.71P_{\text{rot,c}}$, where $P_{\text{rot,c}}$ is the
initial rotational period at the center of the star and is equal to
$\simeq 13M$. At the time the AH is formed, the star has assumed the
shape of a disk which rapidly accretes until no matter is left
outside, as one can see from the last row of panels in
Fig.~\ref{fig:subkerr_dynamics_part1}. Even if we used an ideal-fluid
EOS, we did not see the formation of global shocks during the
collapse.

In Fig.~\ref{fig:subkerr_rhomax} we plot instead the maximum of the
rest-mass density normalized at its initial value for all the three
sub-Kerr models. All the models show the same dynamics with an
exponential increase in the maximum of the rest-mass density and with
the AH forming when $\rho_{\text{max}}$ has reached a value $\sim 2$
times larger than the initial one. As expected, models with a higher
value of $J/M^2$ collapse on a longer timescale. 

In Fig.~\ref{fig:subkerr_restmass} we compare the total rest-mass and
the total angular momentum of all the three models ($\mathtt{A1}$,
$\mathtt{A2}$, $\mathtt{A3}$) normalized to their initial values. As
one can see from this figure, the level of accuracy for these
simulations is very high, with a relative error in the conservation of
rest-mass and angular momentum smaller than $10^{-3}$. Note also that
the angular momentum is computed from the matter sources, outside the
apparent horizon, and hence it is shown to vanish at late times in
Fig.~\ref{fig:subkerr_restmass}, as the matter is dissipated near the
singularity (see discussion in~\cite{baiotti06,Thierfelder10}).

\subsection{\label{ssec:suprakerr}Supra-Kerr Collapse}

Model $\mathtt{B1}$ has $J/M^2\simeq1.1$ and shows very different
dynamics with respect to the sub-Kerr models.  Because in this case we
expected the development of nonaxisymmetric instabilities, we decided
to adopt equatorial and $\pi$-symmetry (this means that we evolved
only the region $\{x>0,z>0\}$ applying a rotational symmetry boundary
condition at $x=0$ and reflection symmetry at $z=0$). Here we report
the results obtained on a grid with boundaries located at $[0,34M]
\times [-34M,34M] \times [0,34M]$ and with a resolution ranging from
$\Delta x^i=0.17 M$ to $\Delta x^i=0.04 M$ with three refinement
levels and with the finest grid covering the entire star. It is worth
remarking that the resolutions adopted here for the supra-Kerr model
are considerably finer than the ones routinely adopted in the
simulation of binary neutron stars (see,
\eg~\cite{Baiotti:2009gk,Giacomazzo:2010,Baiotti:2010,Rezzolla:2011}),
and of a factor $\simeq 4$ larger. As a result, although this scenario
is much simpler to simulate, its computational costs are indeed
equally high.

Because model $\mathtt{B1}$ is a very stable configuration, we had to
enforce its collapse by {artificially} reducing the initial pressure
by $99\%$, as was done by Duez et al.~\cite{duez04}, while smaller
pressure reductions were found to be insufficient to trigger the
collapse. With its pressure support removed, the model immediately
flattens along the $z$-direction and collapses toward the center on
the equatorial plane, producing a strong shock. After a first bounce,
due to the centrifugal barrier produced by the large angular momentum,
a {quasitoroidal structure} forms, which rapidly fragments into four
clumps (see the snapshots from time $t=1.122P_{\text{rot,c}}$ to
$t=2.154P_{\text{rot,c}}$ in Fig.  \ref{fig:suprakerr_dynamics}) whose
formation was observed also in Ref.~\cite{duez04}. We have also
extracted the Fourier modes of the rest-mass density $\rho$, by
computing at $z=0$ and at different cylindrical radii $\varpi\equiv
\sqrt{x^2+y^2}$, the averages
\begin{equation}
k_m(\varpi') \equiv \int_{z=0,\varpi'} \rho
(\varpi'\cos(\phi),\varpi'\sin(\phi)) 
e^{{\rm i} m \phi} d\phi \,.
\end{equation}
The mode power $P_m$ is then simply given by
\begin{equation}
P_m \equiv
\frac{1}{\varpi_{\rm out}-\varpi_{\rm in}}\int_{\varpi_{\rm in}}^{\varpi_{\rm out}}
|k_m(\varpi)| \mathrm{d}\varpi  \,,
\end{equation}
where $\varpi_{\rm in}$ and $\varpi_{\rm out}$ are chosen to cover the
whole domain (for details and an extensive use of this technique, see
also Refs.~\cite{baiotti06b,Manca07}).  

\begin{figure}
  \begin{center}
  \includegraphics[width=0.45\textwidth]{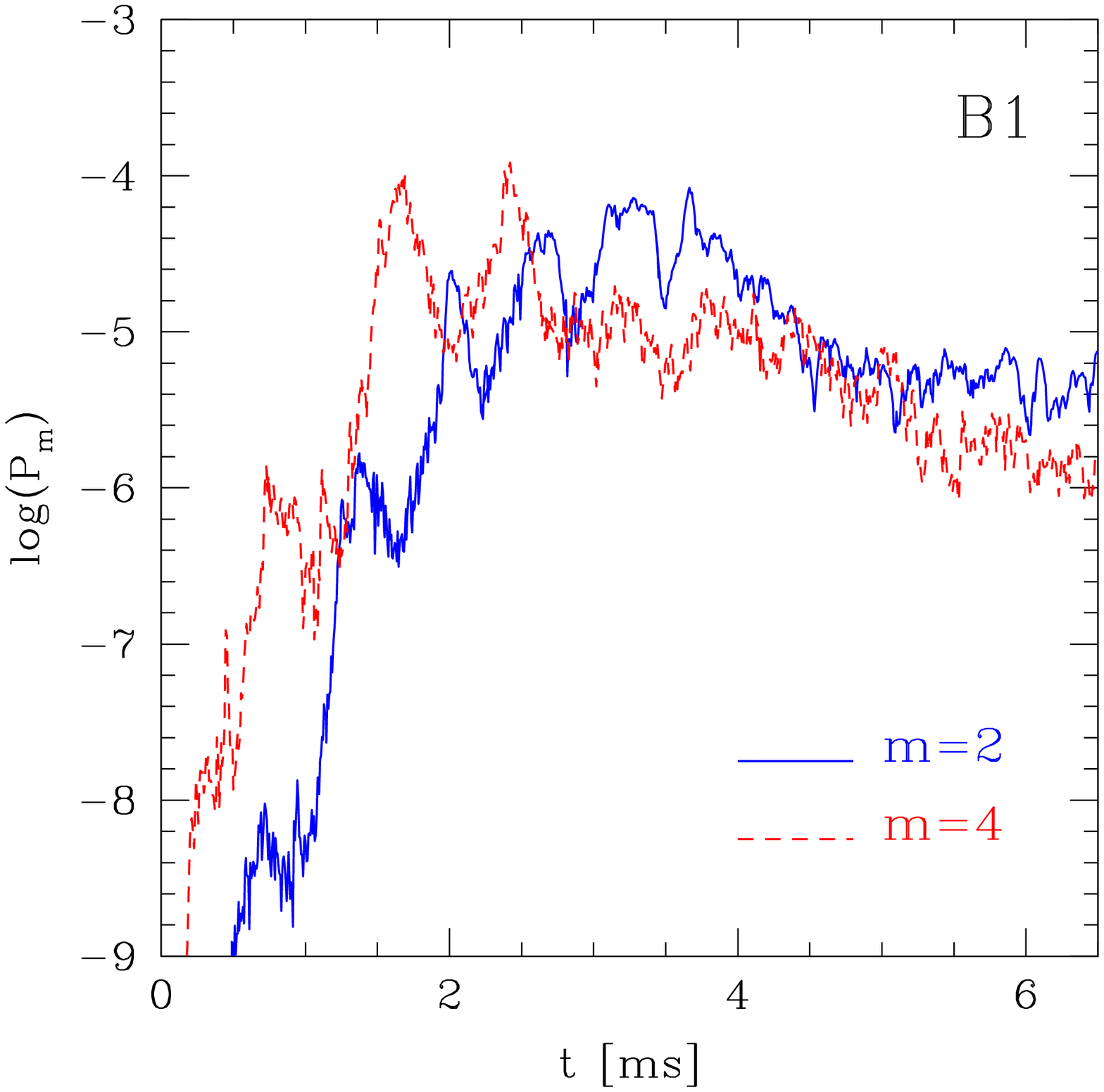}
  \end{center}
  \caption{\label{fig:suprakerr_modepower}Logarithm of the power in
    the $m=2$ (solid line) and $m=4$ (dashed line) Fourier modes as a
    function of time for model $\mathtt{B1}$. The modes not shown in
    this plot are zero during all of the evolution.}
\end{figure}

The presence of an $m=4$ mode at the beginning can be then seen
looking also at the modes' power plotted in
Fig.~\ref{fig:suprakerr_modepower}, where we show the evolution of the
$m=2$ (solid line) and $m=4$ (dashed line) modes, being the other
modes either zero or much smaller during the entire simulation. It is
not clear at the moment whether the fragmentation has to be considered
physical and only triggered by the use of a Cartesian grid, or
entirely due to our Cartesian coordinate system. We note that Truelove
and collaborators~\cite{truelove97} have shown that spurious
fragmentation can occur if the Jeans length is not well resolved,
\textit{i.e.} if the following ``Jeans condition'' is verified,
\begin{equation}
\label{JC}
\frac{\Delta x}{\lambda_J} \gtrsim \frac{1}{4}\,,
\end{equation}
where $\lambda_J$ is the Jeans length and is given by
\begin{equation}
\lambda_J \approx \sqrt{\left(\frac{\pi c_s^2}{\rho}\right)}
\end{equation}
and $c_s$ is the sound speed. Duez et al.~\cite{duez04} estimated the
minimum of the Jeans length to be $\lambda_J\approx 1.3M$ for a model
similar to our model $\mathtt{B1}$ and using a polytropic EOS (for an
ideal-fluid EOS, as the one used in our simulations, the sound speed
is generically larger). In the simulation of Duez et
al.~\cite{duez04}, but also in ours, the value of $\Delta x/\lambda_J$
is indeed found to be smaller than $0.25$, leading to the conclusion
that the fragmentation is physical and due to a genuine
nonaxisymmetric instability. We should remark, however, that in
Ref.~\cite{truelove97} the condition~(\ref{JC}) was indicated as
necessary but not in general sufficient to avoid the formation of
spurious fragmentation. Thus, even if the resolutions used in our
simulation and in the one reported in Ref.~\cite{duez04} satisfy the
Jeans condition, we cannot thus {strictly} exclude that the origin
of this $m=4$ mode is due to the use of a Cartesian grid. Further
investigations with resolutions higher than the ones that could be
afforded here and with different coordinate systems will be necessary
in order to clarify this issue.

\begin{figure}
  \includegraphics[width=0.45\textwidth]{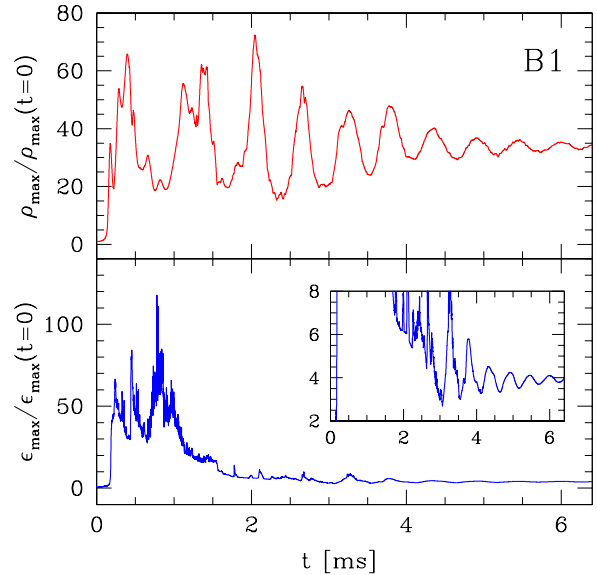}
  \caption{\label{fig:suprakerr_rhomax} The maximum value of rest-mass
    density $\rho$ and of the specific internal $\epsilon$ normalized
    to their initial value and during the evolution of the supra-Kerr
    model $\mathtt{B1}$. Note that the end-configuration is that of
    equilibrium, with larger central density and internal energy. This
    figure should be compared with the corresponding
    Fig.~\ref{fig:subkerr_rhomax}, which refers to the sub-Kerr models
    $\mathtt{A1-A3}$.}
\end{figure}

\begin{figure}
  \begin{center}
  \includegraphics[width=0.45\textwidth]{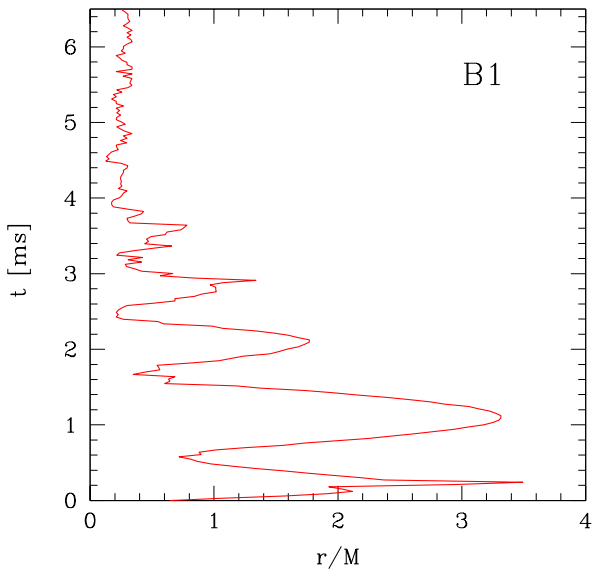}
  \end{center}
  \caption{\label{fig:suprakerr_wordline}Worldline of the maximum of
  the rest-mass density $\rho$ for model $\mathtt{B1}$. The final configuration
  has a maximum located at a radius smaller than the initial one. Note
  also the three bounces.}
\end{figure}

At time $t\approx 2.5 P_{\text{rot,c}}$ the four fragments merge and a
new collapse and bounce follows with the formation of a new
{quasitoroidal structure}. The effects of these bounces on the
maximum of the rest-mass density $\rho$ and specific internal energy
$\epsilon$ are shown in Fig.  \ref{fig:suprakerr_rhomax}.  At
$t\approx 3.0 P_{\text{rot,c}}$ the {quasitoroidal structure
  contracts toward} the center, forming a new configuration which
does not collapse further, but that develops a bar which lasts for
$\approx 2$ ms. At this point, the model approaches a new stable
configuration, as one can easily see from the worldline of the
maximum of the rest-mass density $\rho$
(Fig.~\ref{fig:suprakerr_wordline}) and from its evolution
(Fig.~\ref{fig:suprakerr_rhomax}, top panel). It is also evident from
Fig.~\ref{fig:suprakerr_modepower} that at late times the $m=2$ mode
dominates.

\begin{figure}
  \includegraphics[width=0.45\textwidth]{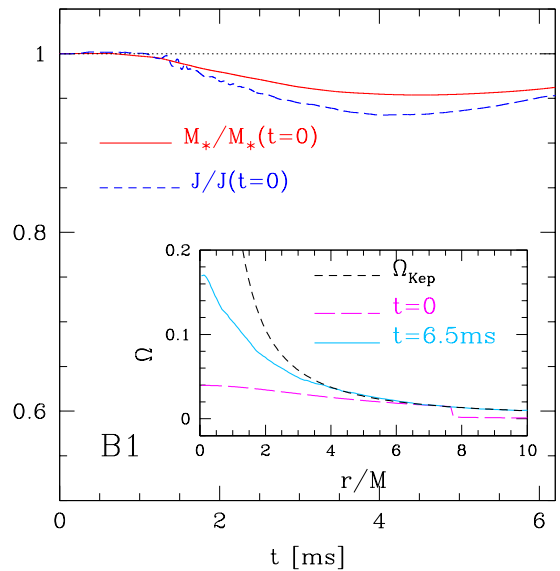}
  \caption{\label{fig:suprakerr_momentum} Total rest-mass $M_*$ and
    total angular momentum $J$ normalized at their initial values for
    the supra-Kerr model $\mathtt{B1}$. The inset shows instead the
    profiles of the angular velocity $\Omega$ at the initial time
    (purple long-dashed line) and at the end of the simulation (light blue
    solid line). For comparison we also show the Keplerian profile
    $\Omega_K \propto r^{-3/2}$ (black short-dashed line).}
\end{figure}

In Fig.~\ref{fig:suprakerr_momentum} we plot the evolution of the
total rest-mass and of the total angular momentum normalized to their
initial values. When we stopped the simulation, $J$ had dropped by
$\sim 10\%$. The loss in the angular momentum cannot be accounted for
in full by the emission of gravitational waves and it is due to the
loss of mass, which is expelled by shocks through the external
boundaries. In the inset of Fig.~\ref{fig:suprakerr_momentum} we plot
instead the profiles of the angular velocity along the $x$-axis,
$\Omega = (\alpha v^y - \beta^y)/x$ at the initial time (purple
long-dashed line) and at the end of the simulation (light blue solid
line), and where $v^j$ is the three-velocity of the fluid as measured
by an Eulerian observer~\cite{whisky}. For comparison we also show the
Keplerian profile $\Omega_K \propto r^{-3/2}$ (black short-dashed
line). It is evident that the collapse leads to an even larger degree
of differential rotation than the initial one and that the new
equilibrium is very close to a Keplerian configuration in its outer
layers. This is clearly the result of having essentially removed
pressure forces, leaving the centrifugal ones the only ones responsible for
the equilibrium.

As a final remark we note that the fact that we were not able to force
this model to collapse to a black hole, even when artificially
reducing the pressure by $99\%$, confirms that supra-Kerr models
cannot directly collapse to a black hole. Furthermore, the evidence
that even when forced to collapse the supra-Kerr model does not
produce a black hole, but rather redistributes its angular momentum to
reach a stable and axisymmetric stellar configuration, provides strong
evidence that cosmic censorship is not violated and that rather
generic conditions for a supra-Kerr progenitor do not lead to a naked
singularity.

%------------------------------------------------------%
\section{\label{sec:waves}Gravitational-wave emission}
%------------------------------------------------------%
We now concentrate on the emission of gravitational waves from the
sub-Kerr and supra-Kerr models, with the aim of comparing our results
with those obtained in Refs.~\cite{baiotti05,Baiotti07} for the
collapse of uniformly rotating neutron stars models.

\subsection{\label{ssec:subkerr_waves}Sub-Kerr Models}

\begin{figure}
   \begin{center}
   \includegraphics[width=0.45\textwidth]{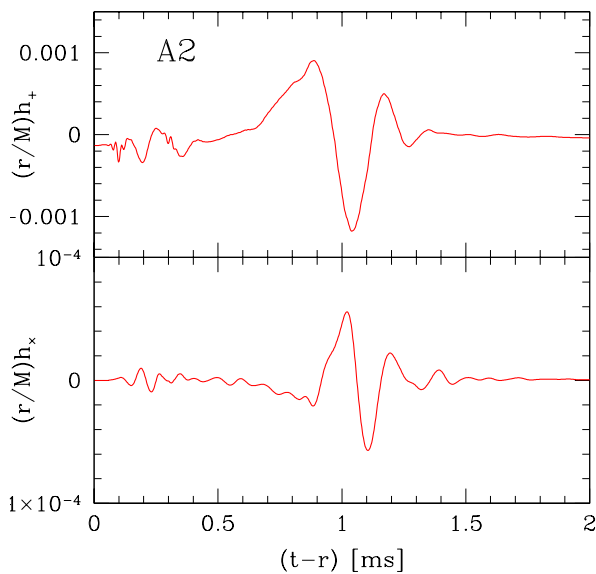}
   \end{center}
   \caption{\label{fig:subkerr_waves_h_all} Gravitational-wave
     amplitudes $h_{+}$ (top panel) and $h_{\times}$ (bottom panel) for
     the collapse of model $\mathtt{A2}$. The waves are computed using
     gauge-invariant perturbations of a Schwarzschild spacetime.}
 \end{figure}

In Fig.~\ref{fig:subkerr_waves_h_all} we show $h_{+}$ and $h_{\times}$
for the $\mathtt{A2}$ sub-Kerr model. We recall that in the case of
axisymmetric collapse to black hole (BH) the signal is expected to be
composed by an initial rapid increase in the amplitude, due to the
initial phase of the collapse, followed by a ring-down phase of the
formed BH~\cite{Cunningham79, Seidel90, Seidel91, baiotti05,
  baiotti06, Baiotti07}. These two phases are very evident in
Fig.~\ref{fig:subkerr_waves_h_all} and the same behavior is shown also
by the other sub-Kerr models. The initial oscillations visible in the
$h_{+}$ for the first $0.5 \mathrm{ms}$ are spurious effects related
to the initial violation of the Hamiltonian constraint introduced by
importing axisymmetric models computed on a spherical grid onto the
Cartesian grid used for the evolution. We have verified indeed that
these oscillations decrease in amplitude with increasing resolution,
while the rest of the signal remains the same.

In Table~\ref{tab:SNR} we report the signal-to-noise ratio for various
second and third-generation gravitational-wave detectors and the
energy emitted through gravitational waves for all the sub-Kerr
models, assuming a galactic source at a distance of $10\,{\rm
  kpc}$. For example, gravitational waves from the collapse of model
$\mathtt{A2}$, with a total energy of about $1.4\times10^{-7}M$, would
arrive at the Virgo detector with a signal-to-noise ratio of $S/N
\approx 2.11$ (with a characteristic amplitude of
$h_c=6.14\times10^{-21}(M/M_{\odot})$ and at a characteristic
frequency $f_c=1367\mathrm{Hz}$). In the case of LIGO, instead, we
obtain $h_c=5.28\times10^{-21}(M/M_{\odot})$ at $f_c=1133\mathrm{Hz}$
with $S/N \approx 1.5$ while for advanced LIGO we have
$h_c=3.81\times10^{-21}(M/M_{\odot})$ at $f_c=809\mathrm{Hz}$ with
$S/N \approx 14.7$. However, when consider third-generation detectors
such as the Einstein Telescope (ET)~\cite{Punturo:2010}, the
signal-to-noise ratio would increase dramatically to $S/N \approx 247$
and the source could then be detected up to a distance of $\sim 1$
Mpc. {Even though detection of individual events is difficult, the
  case of a stochastic background is worth further investigation.}

\begin{figure}
  \begin{center}
  \includegraphics[width=0.45\textwidth]{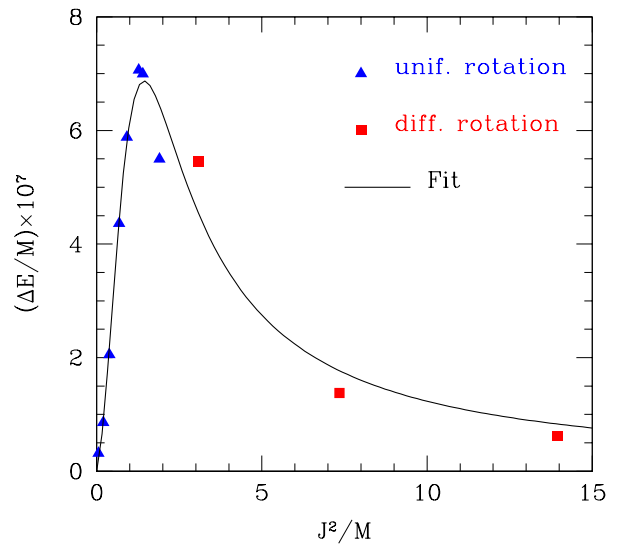}
  \hskip 1.0cm
  \end{center}
  \caption{\label{fig:energy} Energy emitted through gravitational
    waves, normalized to the total initial mass $M$, as a function of
    $J^2/M$. Triangles represent the uniformly rotating neutron star
    models studied in~\cite{Baiotti07}, while the squares refer to the
    differentially rotating models discussed here. The solid line is
    the best fit (see text for details).}
\end{figure}

The gravitational-wave energies computed here are similar to those
obtained for the collapse of uniformly rotating neutron stars, as
reported in~\cite{Baiotti07}. This is to be expected, since even
though the initial quadrupole moment is larger for differentially
rotating models, the collapse proceeds more slowly, due to the higher
values of $J/M^2$ and this reduces the efficiency in energy emission
through gravitational waves. This can also be seen in
Fig.~\ref{fig:energy}, where we plot the total energy emitted by
gravitational waves, $\Delta E/M$, as a function of $J^2/M$, which is
proportional to the initial axisymmetric quadrupole moment. The
triangles represent the uniformly rotating neutron stars discussed
in~\cite{Baiotti07}, the squares represent the differentially rotating
models discussed here, while the solid line is the best fit of the
data with the following analytic expression:
\begin{equation}
\frac{\Delta E}{M} = \frac{\left(J^2/M\right)^{n_1}}
{a_1 \left(J^2/M\right)^{n_2} + a_2} \,,
\end{equation}
where $a_1=(5.17\pm4.37)\times 10^5$, $a_2=(1.11\pm0.57)\times 10^6$,
$n_2=2.63\pm0.53$, and $n_1=1.43\pm0.74$. Clearly, because of the small
statistics, the error in some of the coefficients is rather large, but
this can be compensated by performing additional simulations.

 \begin{table*}
 \caption{\label{tab:SNR}Signal-to-noise ratio (SNR) computed for the
   collapse of sub-Kerr and supra-Kerr models assuming a source at a
   distance of $10\,{\rm kpc}$. The last column shows also the energy
   emitted through gravitational waves which, in the case of the
   sub-Kerr models, is comparable to the one obtained
   in~\cite{Baiotti07} from the collapse of uniformly rotating neutron
   stars.}
 \begin{ruledtabular}
 \begin{tabular}{ccccccc}
   Model          & SNR (Virgo) & SNR (LIGO) & SNR (Adv. Virgo) & SNR (Adv. LIGO) & SNR (ET)  & $\Delta E/M$ \\
   $\mathtt{A1}$  & $1.00$      & $0.60$     & $3.56$           & $4.80$          & $107.41$  & $5.46 \times 10^{-7}$\\
   $\mathtt{A2}$  & $2.11$      & $1.47$     & $13.98$          & $14.67$         & $246.77$  & $1.38 \times 10^{-7}$\\
   $\mathtt{A3}$  & $4.55$      & $3.79$     & $45.70$          & $47.99$         & $594.54$  & $6.24 \times 10^{-8}$\\
   $\mathtt{B1}$  & $16.13$     & $9.39$     & $45.30$          & $71.75$         & $1688.95$ & $7.89 \times 10^{-4}$ \\
 \end{tabular}
 \end{ruledtabular}
 \end{table*}

\subsection{\label{ssec:suprakerr_waves}Supra-Kerr Model}

\begin{figure}
   \begin{center}
   \includegraphics[width=0.45\textwidth]{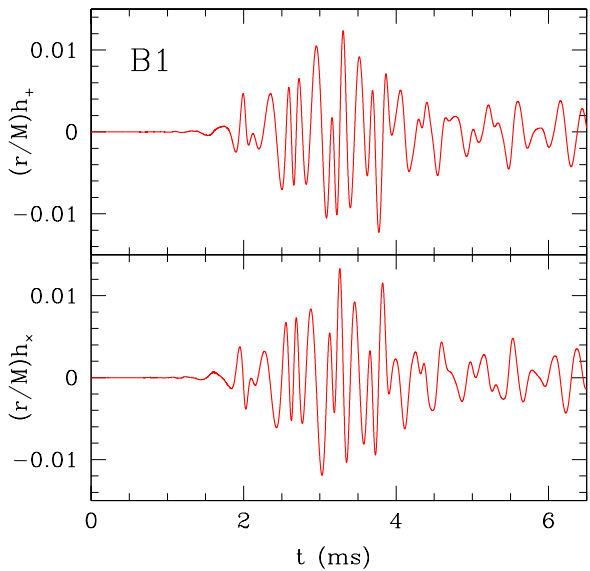}
   \end{center}
   \caption{\label{fig:suprakerr_waves}Gravitational-wave amplitudes
     $h_{+}$ and $h_{\times}$ for the collapse of model
     $\mathtt{B1}$. The waves are computed using the Newtonian
     quadrupole formula.}
 \end{figure}

As anticipated in Sec.~\ref{sec:dynamics}, we have also estimated the
gravitational-wave signal emitted by model $\mathtt{B1}$ using the
standard quadrupole formula~\cite{Nagar2007,Baiotti:2008nf} since in
this case the outer boundary was not located far enough from the
source to allow for the extraction of the signal using the method
described in Sec.~\ref{gwmethod}. In this approximation, the observed
waveform and amplitude for the two polarizations for an observer
situated at large distance $r$ along the $z$-axis are approximately
given by~\cite{baiotti06b}
\begin{eqnarray}
\label{eq:GWquadrupole}
h_{+}     &=& \frac{\ddot{I}^{xx}(t)-\ddot{I}^{yy}(t)}{r}\, , \\
h_{\times} &=& 2\; \frac{\ddot{I}^{xy}(t)}{r} \, ,
\end{eqnarray}
where
\begin{equation}
\label{eq:defQuadrupole}
I^{jk} = \int d^{3}\!x \; D \; x^{j} x^{k}
\end{equation}
is the quadrupole moment of the matter distribution, $D\equiv
\sqrt{\gamma}\rho W$, $\gamma$ is the determinant of the three metric,
and $W$ is the Lorentz factor. The results for the gravitational-wave
amplitudes $h_{+}$ and $h_{\times}$ are finally reported in
Fig.~\ref{fig:suprakerr_waves} and they are 1 order of magnitude
larger than those emitted by the sub-Kerr models. This produces
consequently much larger signal-to-noise ratios as one can see by
looking at the last row of table~\ref{tab:SNR}. As an example, if such
a collapse happened at a distance of $10 \,{\rm kpc}$, we would obtain
in the case of advanced LIGO a signal-to-noise ratio of $S/N \approx
72$, with a characteristic amplitude of
$h_c=1.32\times10^{-19}(M/M_{\odot})$ and at a characteristic
frequency $f_c=2714\mathrm{Hz}$. In the case of ET instead the signal
could be detected for a source at a distance of up to $\approx 10$
Mpc.

\section{\label{sec:diffot_conclusions}Conclusions}

We have presented new results on the collapse of differentially
rotating neutron stars in full general relativity, using simulations
with fixed mesh refinement and in 3D. We have considered two different
classes of initial configurations, consisting of either sub-Kerr
models with $J/M^2<1$, or of supra-Kerr models with $J/M^2>1$. In
particular, we have performed a detailed study of the parameter space
of different equilibrium configurations and, for polytropes which
could represent neutron-star models, we could only find models
unstable to an axisymmetric instability that are sub-Kerr.

When evolving three representative sub-Kerr models with different
degrees of differential rotation we observed in all cases a dynamics
which is very similar to the one already found in the case of uniform
rotation.  In particular, all models remain axisymmetric during
collapse and produce a Kerr black hole. We have also studied the
collapse of a supra-Kerr model when its collapse is triggered by a
$99\%$ depletion of its pressure. In this case, we observed a very
different dynamics, with the formation of nonaxisymmetric
instabilities, the development of a torus and its subsequent
fragmentation in four clumps that merge again, forming a bar and
eventually a stable axisymmetric configuration. While lack of evidence
cannot be taken to exclude the possibility that a naked singularity
can be produced by the collapse of a differentially rotating star, our
results also suggest that cosmic censorship is not violated and that
rather generic conditions for a supra-Kerr progenitor do not lead to a
naked singularity.

For all the models presented here, we have also computed the
gravitational-wave signal, which is comparable to the one from
uniformly rotating neutron stars. In particular, the efficiency in the
emission of gravitational waves for the sub-Kerr models is slightly
smaller than the one of uniformly rotating stars. This is due to the
large angular momentum that differentially rotating neutron stars have
and which increases the time scale on which the collapse
happens. Therefore, even if the initial quadrupole moment is much
larger, the longer time needed for the collapse decreases the overall
efficiency. Furthermore, as for the collapse of uniformly rotating
neutron stars, also in this case the detection of such events would be
possible only for sources located in our Galaxy in the case of the
advanced Virgo and advanced LIGO detectors. However, a
third-generation detector such as ET could instead be able to detect a
source at a distance of $\sim 1$Mpc.

We plan to extend the work presented here by making use of the
axisymmetric code developed in Ref.~\cite{Kellerman08}, which will
allow us to consider a larger number of models and EOSs. We also plan
to investigate the effect that magnetic fields have on these
configurations and especially on the supra-Kerr models by using the
GRMHD version of \texttt{Whisky}~\cite{giacomazzo07}. The magnetic
field can in fact redistribute the angular momentum inside the star
and thus alter the dynamics of the collapse, as already studied in
axisymmetry~\cite{duez06a,duez06b,duez06c} and in
3D~\cite{Giacomazzo:2010}.

\bigskip
%--------------------------------------------------%
%ACKNOWLEDGMENTS
\acknowledgments
This project was initiated during the PhD work of B.G. and it has
benefited over the years of useful discussions and comments from
L. Baiotti, R. De Pietri, I. Hawke, G.~M. Manca, A. Nagar, C.~D. Ott
and E. Schnetter, whom we thank. The numerical computations were
performed on clusters {\it Peyote}, {\it Belladonna} and {\it Damiana}
at the AEI; {\it Albert2} at the Physics Department of the University
of Parma (Parma, Italy); {\it CLX} at CINECA (Bologna, Italy); and
RANGER at TACC through the TERAGRID allocation TG-MCA02N014. This work
was supported in part by the DFG Grant SFB/Transregio~7, by
``CompStar'', a Research Networking Programme of the European Science
Foundation. B.G. acknowledges partial support from NASA Grant No.
NNX09AI75G. N.S.  acknowledges partial support from Grant No. MNiSWN N203
511238.
%--------------------------------------------------%

\bibliography{diffrot}

\end{document}